\documentclass[]{aastex7}
\usepackage{hyperref}
\usepackage{caption}

\begin{document}

\title{Distances of Supernova Remnants Associated with Neutron Stars in the Galaxy}

\correspondingauthor{Shu Wang}
\email{shuwang@nao.cas.cn}

\author{Xiaohan Chen}
\affiliation{CAS Key Laboratory of Optical Astronomy, National Astronomical Observatories, Chinese Academy of Sciences, Beijing 100101, China}
\affiliation{School of Astronomy and Space Science, University of the Chinese Academy of Sciences, Beijing, 100049, China}
\email{chenxh@bao.ac.cn}

\author[0000-0003-4489-9794]{Shu Wang}
\affiliation{CAS Key Laboratory of Optical Astronomy, National Astronomical Observatories, Chinese Academy of Sciences, Beijing 100101, China}
\affiliation{School of Astronomy and Space Science, University of the Chinese Academy of Sciences, Beijing, 100049, China}
\email{shuwang@nao.cas.cn}

\author[0000-0001-7084-0484]{Xiaodian Chen}
\affiliation{CAS Key Laboratory of Optical Astronomy, National Astronomical Observatories, Chinese Academy of Sciences, Beijing 100101, China}
\affiliation{School of Astronomy and Space Science, University of the Chinese Academy of Sciences, Beijing, 100049, China}
\affiliation{Institute for Frontiers in Astronomy and Astrophysics, Beijing Normal University, Beijing 102206, China}
\email{chenxiaodian@nao.cas.cn}

\begin{abstract}

Accurate distance measurements to supernova remnants (SNRs) are essential for determining their physical parameters, such as size, age, explosion energy, and for constraining the properties of associated neutron stars (NSs). We present an extinction–distance method that combines precise \textit{Gaia} DR3 photometry, parallax, and stellar parameters from the SHBoost catalog to homogeneously construct extinction--distance profiles for 44 NS-associated Galactic SNRs. Applying a statistical model, we identify clear extinction jumps along each sightline, corresponding to probable SNR distances. We classify the results into three reliability levels (A, B, and C), primarily based on comparisons with previously reported kinematic distances, supplemented by independent estimates from other methods. Our results show that the majority of reliable distances (17 Level A and 8 Level B) are located within 5 kpc, predominantly in the Local Arm. This study presents an independent and effective method for determining distances to SNRs, particularly for those with small angular sizes or located in the second and third Galactic quadrants.
Although the current method is limited to within 5 kpc due to the precision constraints of \textit{Gaia} parallax and photometry, the upcoming \textit{Gaia} DR4 release, combined with complementary infrared data, will extend its applicability to more distant and heavily obscured SNRs, and help resolve kinematic distance ambiguities.

\end{abstract}



\section{Introduction} 
Supernova remnants (SNRs) are the expanding debris of supernova explosions and often host compact stellar remnants such as neutron stars (NSs) \citep{1985ComAp..11...15S, 1998nspt.conf..401K}. The distance to an SNR is a fundamental and important physical parameter to study its size, luminosity, age, mass, expansion velocity, and the explosion energy of the progenitor supernova. 
Similarly, accurate distance measurements are essential for characterizing the intrinsic properties of associated NSs, such as their luminosity, spin-down energy, and thermal evolution \citep{2003ApJ...593L..89B, 2004Sci...304..536L, 2019LRCA....5....3P}. 
However, determining NSs distances remains challenging. Geometric parallax measurements are often unavailable or have large uncertainties \citep{2012ApJ...755...39V}, and dispersion measure (DM)--based distances rely on Galactic electron density models that carry significant systematic errors \citep{2017ApJ...835...29Y, 2021PASA...38...38P}. These limitations are especially severe for young NSs, whose early evolution is of great astrophysical interest. When a NS is still embedded in its supernova ejecta, direct observation of its birth properties is impossible. Fortunately, NS--SNR pairs offer a promising solution. When a NS is spatially and physically associated with a SNR, the distance to the SNR--often more accessible via methods--can be adopted as the NS distance. In this way, NS--SNR pairs serve as valuable laboratories, where accurate SNR distances help establish more reliable distances for young NSs and thus improve our understanding of their origin and early evolution.

Historically, a variety of methods have been used to estimate distances to SNRs in both the Milky Way \citep{2019JApA...40...36G} and external galaxies \citep{1973ApJ...180..725M}. The most common method is to obtain kinematic distances of SNRs based on their H\,\textsc{I} absorption or CO molecular line emission from the clouds interacting with them \citep{2010ASPC..438..365L, 2011ApJ...727...43S, 2018AJ....155..204R}. While widely adopted, this method suffers from the near/far ambiguity in the inner Galaxy and depends on the assumed Galactic rotation curve, which introduces systematic uncertainties. Moreover, constructing high-quality H\,\textsc{I} absorption spectra remains observationally challenging.
Another approach applicable to shell-type SNRs is the $\Sigma-\rm{D}$ relation, an empirical power-law correlation between the radio surface brightness ($\Sigma$) and the physical diameter (D) of the SNR \citep{1998ApJ...504..761C,2013ApJS..204....4P,2023ApJS..265...53R}. However, this method also has large uncertainties due to significant difference of power-law index and the limitations imposed by observational sensitivity. 
In some cases, SNRs distances are also inferred from their associated objects, including OB associations \citep{1999ApJ...515L..25C} or pulsars \citep{2002astro.ph..7156C}. For particularly nearby SNRs, their distances can be determined by combining proper motion and expansion velocity \citep{2008ApJ...689..231V}.
Each method has its limitations, and consequently many SNR distances remain uncertain. 

In recent years, a promising alternative method, the extinction distance method, has been developed and refined. Supernova is one of the main sources of kinetic energy for the interstellar medium (ISM), stirring up the gas, providing turbulent support to the ISM, and processing interstellar dust through powerful shocks \citep{2005pcim.book.....T}. As a result, SNRs significantly impact their surrounding environment, interacting with nearby dense clouds. When the light from background stars passes through an SNR, the observed stellar extinction increases sharply at the location of the remnant--a phenomenon known as an extinction jump.  By measuring the extinction and distances of a large number of stars, distances of SNRs are estimated by identifying the position of the extinction jump of stars along the SNR sightline \citep{2017MNRAS.472.3924C, 2018ApJ...855...12Z, 2019MNRAS.488.3129Y, 2020ApJ...891..137Z}. In many studies,  red clump stars (RCs) are commonly used as tracers to determine the distances to SNRs \citep{2018ApJS..238...35S,2019RAA....19...92S,2020AA...639A..72W} and molecular clouds (MCs) \citep{2020MNRAS.496.4637C}. However, this method is limited by the availability of a sufficient number of RCs or stars with both precise distances and extinction measurements especially for SNRs with very small size.

The \textit{Gaia} DR3 catalog provides photometric, spectroscopic, and astrometric data for over 1.8 billion sources. Even for SNRs with sub-arcminute angular sizes, numerous stars with both parallax and photometric measurements are available along the line of sight. Moreover, stellar parameters such as effective temperature ($T_{\rm eff}$), surface gravity ($\log~g$), and metallicity have been reliably derived from BP/RP spectra with relatively high precision \citep{2023ApJS..267....8A,2023MNRAS.524.1855Z,2024MNRAS.52710937Y,2024A&A...691A..98K}. These allow accurate extinction and distance estimates for a wide range of stars. By identifying extinction jumps, this approach can extend distance estimates to more SNRs, including those with small angular sizes.

In this work, we use photometric and astrometric data from \textit{Gaia} DR3, combine with stellar parameters from \citet{2024A&A...691A..98K}, to homogeneously determine the distances of 44 SNRs that are associated with NSs \citep{2022MNRAS.514.4606I}. Our extinction-based distance method based on accurate \textit{Gaia} parallax and reliable stellar parameters, combining with widely adopted kinematic distance estimates, offers an independent method to validate or cross-check SNRs distances. 
In Section \ref{sec:data}, we describe the astrometric, photometric data, stellar parameters used in this work, and SNRs samples. Section \ref{sec:method} outlines our methodology, including the derivation of extinction--distance distributions based on corrected \textit{Gaia} parallax and derived color excess, the construction of extinction--distance model along each line of sight to estimate distances of SNRs, and additional treatments for special cases. The derived distances for individual SNRs are represented in Section \ref{sec:res}, together with an analysis of the influence of different fitting approaches on the distance estimates. Finally, in Section~\ref{sec:summary}, we summarize the main results of this work.

\section{Data and sample} \label{sec:data}
\subsection{Gaia} \label{sec:gaia}
\textit{Gaia} DR3 contains over 1.8 billion sources with photometric, spectroscopic and astrometric data. In this work, stellar distances are derived from \textit{Gaia} parallaxes. Since the parallax zero-point in \textit{Gaia} DR3 is known to be affected by stellar magnitude, color, and spatial position, we determined the corrected parallax ($\varpi_{\rm corr}$) using the code provided by \citet{2021A&A...649A...4L}. To ensure reliable distance estimates, we selected stars with parallax $\varpi>0$ and required the fraction parallax uncertainty to be less than $20\%$. In addition, stars with $G$-band magnitudes fainter than 17 were excluded. We further removed objects with ${\rm RUWE} > 1.4$ to reduce the possible influence on parallaxes by photocentric biases \citep{2022A&A...667A..74A}. The richness and precision of the \textit{Gaia} dataset provide an unprecedented number of stars with accurate distance estimates, enabling us to construct extinction–distance profiles even for SNRs with angular sizes as small as a few arcminutes.

\subsection{Stellar Parameters}
\citet{2024A&A...691A..98K} derived stellar parameters from \textit{Gaia} DR3 XP spectra using a gradient-boosted random forest regressor (XGBoost). The model was trained on a dataset of approximately 7 million stars from high-quality parameters provided by the \texttt{StarHorse} code and major spectroscopic surveys. This approach enabled the estimation of reliable stellar parameters for about 217 million stars, including white dwarfs and hot stars that are often underrepresented in other surveys. The resulting SHBoost catalog provides estimates of effective temperature ($T_{\rm eff}$), surface gravity ($\log~g$), metallicity ([M/H]), and stellar mass, along with robust uncertainty estimates. 

We used the SHBoost catalog to obtain stellar parameters ($T_{\rm eff}$, $\log g$, and $\rm [M/H]$) for stars along the line of sight toward each SNR. To ensure parameter reliability, we applied quality cuts based on catalog-provided flags: uncertainties in $\log g$ and [M/H] were required to be below 0.3, and the $T_{\rm eff}$ error below 0.1. We further limited the sample to stars with $T_{\rm eff}$ between 4000 and 8000 K and $\log~g$ between 1 and 5, corresponding to the well-calibrated regime of the SHBoost model \citep[see][]{2024A&A...691A..98K}. Outside this range, especially at $T_{\rm eff} \lesssim 4000$ K and $\gtrsim 8000$ K, the model suffers from reduced generalization and increased uncertainties. We therefore restrict our analysis to this reliable parameter space.

\subsection{SNR Sample}
The SNR sample used in this work is based on the 58 NS--SNR pairs compiled by \citet{2022MNRAS.514.4606I}. 
For each of these SNRs, we defined a corresponding SNR region to select star samples located along the line of sight. We adopted the central coordinates and angular sizes provided in the SNR catalogs of \citet{2019JApA...40...36G} and \citet{2012AdSpR..49.1313F}, which include basic properties for over 300 SNRs. 
Four of the 58 NS--SNR pairs are classified as uncertain in the catalog of \citet{2012AdSpR..49.1313F} and lack reference radio sizes. We excluded these from our sample and remained 54 NS–SNR pairs.  
Based on the center positions and angular radii, we defined spatial region for each SNR and queried the \textit{Gaia} DR3 catalog to extract stars located within each SNR region.
Some SNRs exhibit elliptical projected morphologies due to their irregular morphologies. For these cases, we adopted circular regions with radii equal to their semi-minor axes to define the selection area. This conservative choice minimizes contamination from nearby high-extinction regions, such as MCs and star-forming regions. Then, we cross-matched the selected \textit{Gaia} sources with the SHBoost catalog \citep{2024A&A...691A..98K} to obtain the parallax, $G_{\rm BP}$ and $G_{\rm RP}$ magnitudes, as well as stellar parameters ($T_{\rm eff}$, $\log~g$ and $\rm [M/H]$) for each star.

Our goal is to estimate distances of SNRs with sizes as small as possible. However, our method requires a sufficient number of background stars to construct a reliable extinction--distance profile. We found that SNRs with angular diameters smaller than 5$\arcmin$ typically contain fewer than 15 available stars, which is insufficient for robust distance determination. Consequently, we excluded these 10 SNRs and retained a final sample of 44 SNRs for further analysis.

\section{Method} \label{sec:method}
\subsection{Extinction--Distance Profiles}
To determine the distance to a given SNR, we analyze the variation of extinction with distance along its line of sight, thereby constructing an extinction--distance profile and identifying the location of extinction jumps through statistical modeling. 
A critical step in this process is to estimate the extinction for each star along the SNR sightline, typically expressed as the color excess $E(G_{\rm BP}-G_{\rm RP})$. This color excess is calculated by subtracting the intrinsic color index from the observed color index. Since the observed color index can be directly obtained from \textit{Gaia} photometry, the primary task is to determine the intrinsic color index for each star.

We adopted the ``blue-edge'' method, a widely used approach to estimate stellar intrinsic color indices \citep{2014ApJ...788L..12W, 2019ApJ...877..116W, 2023ApJ...956...26L, 2023ApJ...946...43W, 2025ApJ...982...77D}. This method establishes empirical relations between $T_{\rm eff}$ and intrinsic colors for a given luminosity class. \citet{2024ApJ...974..138Z} recently improved its generalization by training an XGBoost model on over one million low-reddening stars, using \textit{Gaia} DR3 GSP-Phot parameters ($T_{\rm eff}$, $\log~g$, [M/H]) as input. The model predicts intrinsic color indices in both \textit{Gaia} and 2MASS bands, including $(G_{\rm BP}-G_{\rm RP})_0$, $(G_{\rm BP}-K_{\rm S})_0$, and $(J-K_{\rm S})_0$.

In this work, we used the SHBoost catalog \citep{2024A&A...691A..98K} to obtain stellar parameters, and applied the XGBoost model to estimate $(G_{\rm BP}-G_{\rm RP})_0$ for stars along each SNR sightline. The color excess was then calculated as $E(G_{\rm BP}-G_{\rm RP}) = (G_{\rm BP} - G_{\rm RP}) - (G_{\rm BP}-G_{\rm RP})_0$. As the model does not provide individual uncertainties, we adopted a fixed intrinsic color uncertainty of 0.014~mag, following \citet{2024ApJ...974..138Z}, and propagated photometric errors in $G_{\rm BP}$ and $G_{\rm RP}$ to obtain the total uncertainty in color excess. Finally, combining color excesses with distances derived from corrected \textit{Gaia} parallaxes (Section~\ref{sec:gaia}), we constructed extinction--distance profiles for each sightline, which were used to identify extinction jumps and estimate SNRs distances.

\subsection{Distance Determination via Extinction--Distance Modeling}
We determine the distance to each SNR by identifying a significant extinction jump in its extinction--distance profile. Such a jump appears as a sharp increase in extinction over a relatively narrow distance range, beyond which the extinction in the SNR direction exceeds that of the surrounding diffuse ISM. In many cases, this manifests as a distinct break or steepening in the extinction versus distance curve.

To quantify the extinction jump, we adopted the extinction--distance model from \citet{2017MNRAS.472.3924C}. 
This model assumes that the total extinction along the line of sight is primarily contributed by two components: diffuse interstellar dust and the dust associated with the SNR. Accordingly, the total extinction at distance $d$ is expressed as
\begin{equation}
A(d)=A_{0}(d)+A_{1}(d), 
\end{equation}
where $A_{0}(d)$ denotes the extinction contributed by the diffuse interstellar dust, and $A_{1}(d)$ represents the extinction associated with the SNR. 
The diffuse component is modeled as 
\begin{equation}\label{equ1}
A_{0}(d)=a\times{d}+b\times{\sqrt{d}},
\end{equation}
This means that interstellar extinction increases rapidly at close distances due to the influence of inhomogeneous local medium, while at larger distances, it increases more gradually as a result of the cumulative effect of a more uniform ISM. An exponential form was also considered for the ISM model, but it failed to yield satisfactory results. A detailed discussion is provided in Section~\ref{sec:different ISM}.
The SNR-associated extinction is modeled as 
\begin{equation}
A_{1}(d)=\frac{\delta{A}}{2}\times{[1+erf(\frac{d-d_{0}}{\sqrt{2}\delta{D}})]}, 
\end{equation}
where $\delta{A}$ is the amplitude of the extinction jump, $\delta{D}$ is the diameter of the SNR, and $d_{0}$ is the distance to the center of the SNR. The parameters $a$, $b$, $\delta A$ and $d_{0}$ are fitting coefficients.

We performed a Markov Chain Monte Carlo (MCMC) procedure to fit the extinction--distance profiles and derive the best-fit model parameters ($\delta{A}$, $a$, $b$, $d_{0}$). 
The parallax uncertainties were propagated into the color excess uncertainties and incorporated into the likelihood function used during MCMC sampling. This ensures that both astrometric and photometric uncertainties are fully accounted for, yielding statistically robust distance estimates.
In cases where the MCMC posterior distributions exhibited multiple distinct peaks, suggesting the presence of two or more extinction jumps along the line of sight, we fitted the extinction--distance profile using multiple jump components.

Several SNRs (e.g. G6.4-0.1) exhibit nearly constant color excess values ($E(G_{\rm BP}-G_{\rm RP})\approx 0.2-0.5$ mag) at distances less than $\sim$1 kpc, indicating substantial foreground extinction. 
In these cases, the standard form of $A_{0}(d)$ (Equation~\ref{equ1}) could not provide a good fit. We therefore fitted a separate linear extinction model to stars within $\sim 1$ kpc to model the foreground, and subsequently fixed this component during MCMC fitting of the extinction jump parameters ($d_0$, $\delta A$). 

Additionally, 25 SNRs in our sample have angular diameters larger than 20$\arcmin$ and contain a large number of stars (ranging from 2,000 to 70,000, depending on their size), which complicates the visual identification of how extinction varies with distance in the extinction--distance profiles. To address this, we applied a binning scheme in distance space, averaging the color excesses and distances of stars in 10 pc intervals. This approach significantly improves the visibility of extinction jumps and facilitates more stable modeling along heavily populated sightlines. The effectiveness and potential bias introduced by this binning scheme are quantitatively evaluated in Section~\ref{sec:binning_effect}. 
For the remaining 19 SNRs with intermediate angular diameters (5-20$\arcmin$), we retained the original, unbinned stellar samples for analysis.

\section{Results and Discussion} \label{sec:res}
\subsection{The SNR Distances} \label{sec:SNRdist}
Using the extinction--distance model, we successfully derived extinction profiles for all 44 selected SNRs. Figure \ref{fig:1} presents the color excess $E(G_{\rm BP}-G_{\rm RP})$ as a function of distance for these SNRs, as well as the best-fit extinction profiles, including contributions from both the diffuse ISM and the SNR-associated extinction. 
The derived distances, angular diameters, and other fundamental properties of the SNRs are listed in Table \ref{tab:table1}. The listed distance uncertainties correspond to the 1$\sigma$ statistical errors from MCMC sampling procedure. However, these values may underestimate the true errors, as they do not fully account for potential systematic errors arising from stellar distance measurements and extinction estimates.
Among the studied SNRs, 22 SNRs exhibit a single extinction jump along their lines of sight. 21 SNRs show evidence of two extinction jumps, while one source, G54.4$-$0.3, presents three notable extinction jumps.

To evaluate the accuracy and reliability of our extinction-based distance estimates, we primarily compared them with previously reported kinematic distances. This is because the kinematic method is widely adopted for estimating distances to Galactic SNRs, and when well-constrained and supported by independent diagnostics, it is generally regarded as a robust and reliable reference. In addition, we incorporated complementary distance estimates from other independent methods, including those based on RCs, the $\Sigma-\rm{D}$ relation, and X-ray absorption measurements.
Our method, based on extinction-distance profiles derived from precise \textit{Gaia} parallaxes, provides an independent and systematic approach for estimating distances to Galactic SNRs. It is particularly useful for SNRs with uncertain or poorly constrained distances. When combined with kinematic estimates, our method enables effective cross-validation and improves the reliability of distance measurements. Based on comparisons with previous studies, we classified the distances of 44 SNRs into three reliability levels. In particular, for SNRs with multiple extinction jumps, we performed detailed case-by-case analyses (see Section~\ref{sec:compare}) to determine the most reasonable distances. 

\begin{itemize}
\item Level A: SNR with distance estimates consistent with reported kinematic distance within 30\%. It contains two sublevels: 1) Level $\rm A^{+}$: SNRs exhibiting a single clear extinction jump; 2) Level $\rm A^{-}$: SNRs showing two extinction jumps, where the adopted distance is the one that consistent with the kinematic estimate.
\item Level B: SNRs with distances differing by less than 30\% from those estimated by alterative  methods, such as the $\Sigma-\rm D$ relations, RCs, or associated objects. Based on the number of extinction jumps, Level B is also subdivided into two sublevels: 1) Level $\rm B^{+}$: SNRs with a single extinction jump; 2) Level $\rm B^{-}$ SNRs with two extinction jumps.
\item Level C: SNRs falling into one of three subcategories: 1) no detectable extinction jump;
2) weak or unclear extinction jump, even if the derived distance agrees with other estimates; 
3) lack of background stars beyond $\sim$5 kpc, limiting detection of distant extinction jumps.
\end{itemize}

According to this classification, our final results include 17 Level A (10 Level $\rm A^{+}$, 7 Level $\rm A^{-}$), 8 Level B (7 Level $\rm B^{+}$, 1 Level $\rm B^{-}$), and 19 Level C SNRs. These classification levels and corresponding reference distances from the literature are listed in Table~\ref{tab:table1}. 
For Level C, two SNRs (G296.5+10.0 and G263.9$-$3.3) fall into the first subcategory with no significant extinction jump detected. Specifically, stars towards G296.5+10.0 exhibit uniformly low extinction values, with $E(G_{\rm BP}-G_{\rm RP})\lesssim0.35$ mag, suggesting these stars are likely all foreground stars. For G263.9-3.3, previous studies suggested a very nearby distance of 0.23$-$0.25 kpc \citep{1999ApJ...515L..25C,2003ApJ...596.1137D}. However, we did not detect a corresponding extinction jump. A detailed discussion is presented in Section \ref{sec:compare}. 
Two SNRs, G33.6+0.1 and G35.6$-$0.4, are categorized into the second subcategory. While their extinction-based distances are broadly consistent with kinematic estimates, the associated extinction jumps rely on only a small number of stars and lack statistical robustness (see the first and third panels in the third row of Figure~\ref{fig:1}).
The remaining 15 Level C SNRs fall into the third subcategory. All are located in the first and fourth Galactic quadrants and lack extinction jumps at their previously reported kinematic distances, which typically exceed 5 kpc---a range where \textit{Gaia} parallaxes become less reliable. Moreover, the ambiguity inherent in the kinematic distances for these regions further complicates direct comparisons with extinction-based estimates. This limitation is discussed in greater detail in Section~\ref{sec:Spatial Distribution}.

\begin{figure}[h]
\centering
\includegraphics[scale=0.7]{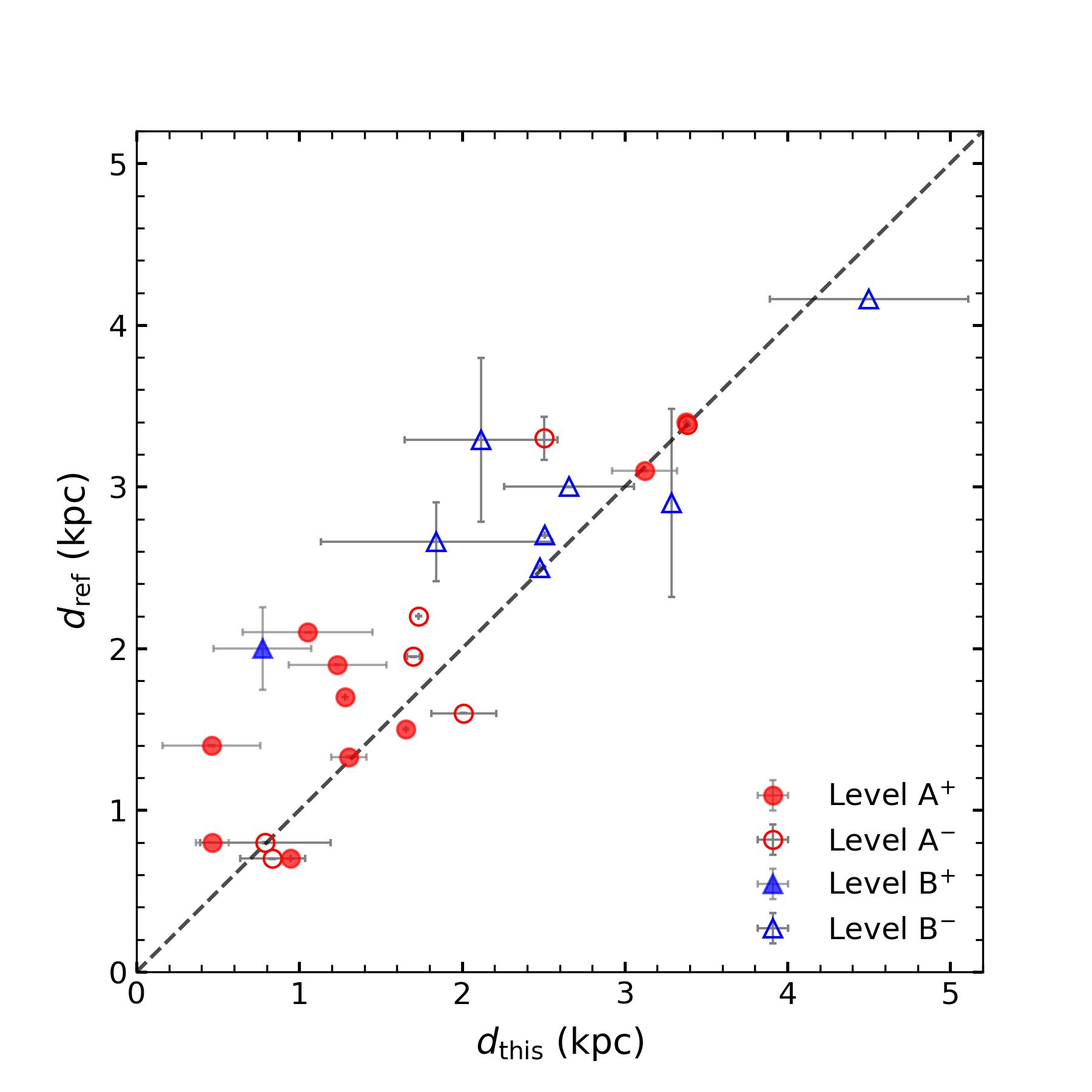}
\figurenum{2}
\caption{Comparison between distances derived in this work (extinction-based distances $d_{\rm this}$) and those reported in the literature ($d_{\rm ref}$). Red circles and blue triangles represent SNRs classified as Level A (consistent with kinematic distances) and Level B (consistent with other independent methods), respectively. Filled symbols (A$^+$/B$^+$) indicate SNRs exhibiting a single clear extinction jump, while open symbols (A$^-$/B$^-$) denote cases with two extinction jumps, where the adopted distance is consistent with the corresponding reference. 
The black dashed line indicates the one-to-one correspondence.}
\label{fig:5}
\end{figure}

For Level A and Level B SNRs, we also present a comparison between our extinction-based distances and previous results in Figure~\ref{fig:5}. The overall agreement is generally good. However, in most cases, the kinematic distances tend to be slightly larger than our extinction-based estimates. This discrepancy may reflect the influence of non-circular motions or local dynamical effects that introduce systematic uncertainties into kinematic distances \citep{2015AJ....150..147F}.

\subsection{Complex Sightlines: Multiple Extinction Jumps}\label{sec:compare}

This section presents a detailed case-by-case evaluation of 21 SNRs that exhibit two significant extinction jumps, as well as one SNR (G54.4$-$0.3) that displays three distinct jumps. The presence of multiple extinction jumps along a single sightline indicates the existence of several dusty clouds, which complicates the identification of the jump associated with the SNR itself. To address this ambiguity, we systematically compare our extinction-based distances with previously reported kinematic distances derived from H\,\textsc{I} and CO observations, as well as other independent estimates based on RCs and X-ray absorption measurements. Based on this comparison, we determine the most probable distance for each SNR and assign a reliability level (A, B, or C) following the criteria outlined in Section~\ref{sec:SNRdist}. This comprehensive approach allows for a robust evaluation of our extinction--distance method, particularly for SNRs whose distances remain uncertain or debated in the literature. The detailed case-by-case discussion is presented below.

\begin{enumerate}
    \item G23.3-0.3 (W41) is an asymmetric shell-type SNR. \citet{2008AJ....135..167L} used H\,\textsc{I} absorption spectra and the \(^{13}\text{CO}\) emission spectra to estimate a distance range of 3.9$-$4.5 kpc, which supports that W41 is physically associated with the MCs at a radial velocity of 78 \textrm{km~s$^{-1}$}. \citet{2018AJ....155..204R} adopted a radial velocity of 78.51 \textrm{km~s$^{-1}$} and revised the distance to $4.8\pm0.2$ kpc base on H\,\textsc{I} and \(^{13}\text{CO}\) observations. However, a smaller distance of $3.38\pm0.26$ kpc was estimated using RCs as distance tracers \citep{2020AA...639A..72W}. In addition, \citet{2023ApJS..268...61Z} inferred that the remnant may be associated with either the +56.6 or +74.1 \textrm{km~s$^{-1}$} MCs, corresponding to distances of $3.5\pm0.4$ kpc or $4.8\pm0.6$ kpc, respectively. This implies that the RC-based result of \citet{2020AA...639A..72W} may trace the component at +56.6 \textrm{km~s$^{-1}$}. In addition, \citet{2013ApJS..204....4P} reported a distance of 2.7 kpc based on the $\Sigma-\rm{D}$ relation. These diverse estimates highlight ongoing uncertainty and debate regarding the true distance to W41, with different tracers yielding results that span more than 1 kpc. Our extinction-based result yields a far distance of 2.508 kpc, which is consistent with the $\Sigma-\rm{D}$ estimate. Therefore, we adopt 2.508 kpc as the distance of W41 and classify it as Level B. 
    
    \item G33.6+0.1 is a shell-type SNR located in a complex region, also known as Kes 79 and HC13. \citet{1975A&A....45..239C} presented an H\,\textsc{I} absorption spectrum for positive velocities and suggested a lower limit distance of 7 kpc. \citet{2016ApJ...816....1K} confirmed a broad line detection and reported a consistent distance of 7.1 kpc. However, \citet{2018AJ....155..204R} revised the distance to 3.5 kpc based on new evidence from the H\,\textsc{I} and \(^{13}\text{CO}\) observations. In our analysis, we identified two extinction jumps at 0.551 and 2.144 kpc. Due to \textit{Gaia} parallax limitations, no stars beyond 2.5 kpc are available in this field. We adopt 2.144 kpc as the distance of G33.6+0.1 based on an unclear extinction jump, and classify it as Level C.
    
    \item G34.7$-$0.4 (W44) is a bright remnant with an elongated and distorted shell. \citet{1972ApJS...24...49R} used H\,\textsc{I} absorption spectra and determined the distance to be 3 kpc. Subsequent H\,\textsc{I} and \(^{13}\text{CO}\) studies yielded similar kinematic estimates ranging from 2.5 to 3.0 kpc \citep{1975A&A....45..239C,1999ApJ...524..179C,2018AJ....155..204R}. Additionally, RCs analysis by \citet{2020AA...639A..72W} also reported a distance of $2.66\pm0.71$ kpc. 
    In our work, we identified two extinction jumps at 0.585 and 1.843 kpc. Considering distance uncertainty, the larger 1.843 kpc is consistent with \citet{2020AA...639A..72W}. So we adopt 1.843 kpc as the distance to G34.7$-$0.4 and classify it as Level B.
    
    \item G35.6$-$0.4 is identified as an SNR by \citet{2009MNRAS.399..177G} from radio and infrared survey observations. It is located in a region rich of diffuse Galactic emission and shows a shell morphology elongated in the northwest-southeast direction. \citet{2013ApJ...775...95Z} estimated its distance to be $3.6 \pm 0.4$ kpc using H\,\textsc{I} and \(^{13}\text{CO}\) spectra, later revised to $3.8 \pm 0.3$ kpc by \citet{2018AJ....155..204R} based on new H\,\textsc{I} and \(^{13}\text{CO}\) observations. 
    In our analysis, a few stars resulting in an unclear extinction jump at 1.509 kpc. Similar to G33.6+0.1, we adopt 1.590 kpc as the distance and assign this SNR to Level C.
    
    \item G53.4+0.0 is identified as an SNR by \citet{2018ApJ...860..133D} through radio, infrared, and X-ray observations. They applied the Galactic spiral arm model and the X-ray absorption, and determined that the SNR is located in the Sagittarius--Carina Arm at a distance of 7.5 kpc. However, we obtained two distances (0.413 and 1.227 kpc) that seem to correspond to foreground MCs. We therefore classify this SNR as Level C. 

    \item G54.1+0.3 is a possible composite-type remnant surrounding the Crab-like PWN G54.1+0.3, located near the H\,\textsc{II} region G54.09$-$0.06. Using H\,\textsc{I} absorption spectra, \citet{2008AJ....136.1477L} estimated a distance of 4.5-9 kpc to the SNR. They suggested that the morphological association of the PWN with a MC at 53 \textrm{km~s$^{-1}$} yields a likely distance of 6.2 kpc to the SNR. \citet{2018AJ....155..204R} derived a distance of 4.9 kpc based on H\,\textsc{I} and \(^{13}\text{CO}\) observations. 
    Same as G53.4+0.0, the two distances (0.62 and 1.405 kpc) estimated in our work likely correspond to foreground MCs. This SNR is therefore classified as Level C. 
    
    \item G54.4$-$0.3 (HC40) is an SNR with an incomplete shell, located in a complex region containing numerous H\,\textsc{II} regions \citep{1985AJ.....90.1224C}. \citet{1992A&AS...96....1J} suggested a distance of 3 kpc based on CO interactions, while the $\Sigma-\rm{D}$ relation yielded and revised estimates of 3.3 kpc and 2.5 kpc \citep{1998ApJ...504..761C,2013ApJS..204....4P}. However, A larger distance of $6.6 \pm 0.6$ kpc was later proposed by \citet{2017ApJ...843..119R} based on H\,\textsc{I} and \(^{13}\text{CO}\) spectra with a central \(^{13}\text{CO}\) velocity of 36.66 \textrm{km~s$^{-1}$}. We identify three extinction jumps for this SNR at 0.545, 1.262, and 2.477 kpc. The most distant of these, 2.477 kpc, is more consistent with previous estimates. The nearer jumps are likely caused by foreground MCs. Therefore, we adopt 2.477 kpc as the distance to G54.4$-$0.3 and classify it as Level B.
    
    \item G65.1+0.6 is a low surface brightness SNR that has been rarely studied. Based on H\,\textsc{I} observations, \citet{2006AA...455.1053T} identified structures associated with the remnant in the radial velocity range of $-20$ to $-26$ \textrm{km~s$^{-1}$} and suggested a kinematic distance of 9.2 kpc. This estimate is significantly larger than our far distance of 4.5 kpc. 
     However, a distance of $4.16 \pm 0.61$ kpc estimated from RC stars by \citet{2020AA...639A..72W} agrees with our result within uncertainties. We therefore adopt 4.5 kpc as the distance of SNR and classify it into Level B. Notably, the H\,\textsc{I} data used by \citet{2006AA...455.1053T} had a velocity resolution of 1.32 \textrm{km~s$^{-1}$}, which may have limited their ability to resolve finer velocity components. Higher-resolution H\,\textsc{I} and CO observations are needed to refine the distance estimate. 
    
    \item G89.0+4.7 (HB~21) is an old SNR with an irregular radio shell. A distance of $1.9^{+0.3}_{-0.2}$ kpc was derived by \citet{2018ApJS..238...35S} using RC stars, while \citet{2020ApJ...891..137Z} derived a distance of $2.3\pm0.3$ kpc. Both applied extinction-jump methods but differed in stellar selection and distance calculation approaches.
    Compared with these works, our analysis includes more stars within the 1.5-2.5 kpc range, allowing for a more accurate distance estimate of 2.01 kpc. Furthermore, this value is also in good agreement with the recent determination of $1.6\pm0.2$ by \citet{2023ApJS..268...61Z}, which assumes that the SNR is associated with a MC. We therefore adopt 2.01 kpc as the distance to G89.0+4.7 and classify it as Level A. The closer extinction jump at 0.369 kpc is likely attributable to the Local Arm or foreground MC.
    
    \item G116.9+0.2 (CTB~1), has been studied as part of a group of SNRs (including G114.3+0.3 and G116.5+1.1) located within a large H\,\textsc{I} shell in the Perseus Arm at a distance of 4.2 kpc, as proposed by \citet{1986ApJ...303..465F}. This study also estimated a distance of 3.4 kpc for G116.9+0.2 by comparing the velocity of the Perseus Arm center with the distances of nearby H\,\textsc{II} regions. However, \citet{2004ApJ...616..247Y} argued that these SNRs lie in the Local Arm, estimating 1.6 kpc for G116.9$+$0.2 based on H\,\textsc{I} observations.
    We obtained two extinction jumps for G116.9+0.2, at 0.412 and 3.383 kpc. The nearer jump likely arises from the Local Arm or foreground cloud, while the far distance is roughly consistent with the result of \citet{1986ApJ...303..465F}. 
    \citet{2020ApJ...891..137Z} estimated a distance of $4.3\pm0.2$ kpc and assigned this SNR to Level B due to a lack of stars in the 1.5-3.5 kpc range. In contrast, our study includes more stars in this range, yielding a clearer extinction jump. We therefore adopt 3.383 kpc as the distance to G116.9+0.2 and classify it to Level A. 

    \item G132.7+1.3 (HB~3) is located adjacent to the H\,\textsc{II} region/MC complex W3 and has been proposed to be physically associated with it \citep{1987AJ.....94..111L}. This association was later confirmed by \citet{2016ApJ...833....4Z}. They used CO observation to identify HB~3 interacts with the $-45$ \textrm{km~s$^{-1}$} MC, corresponding to a distance of $1.95\pm0.04$ kpc. Recently, \citet{2023ApJS..268...61Z} further validated the associated with the $-44.0$ \textrm{km~s$^{-1}$} MC component and derived a distance of $ 1.96\pm0.04$ kpc. Our extinction-based estimate yields a far distance of 1.701 kpc, representing a $\sim$13\% difference from the previous results. So we adopt 1.701 kpc as the distance to HB~3 and classify it as Level A. 
     
    \item G160.9+2.6 (HB~9). \citet{2007AA...461.1013L} determined a distance of $0.8\pm0.4$ kpc, which is roughly consistent with our closer extinction jump at 0.791 kpc. Additionally, \citet{2019MNRAS.488.3129Y} and \citet{2023ApJS..268...61Z} also suggested associations with MCs at distances of 0.6 kpc and $0.5\pm0.2$ kpc, respectively. Considering distance uncertainty, all previous estimates are approximately consistent with our result of 0.791 kpc. Therefore, we adopt 0.791 kpc as the distance to HB~9 and classify it as Level A. The more distant extinction jump at 1.489 kpc is likely attributable to the Perseus Arm or background MCs.
    
    \item G260.4$-$3.4 (Puppis~A, MSH~08-44), SNR G263.9$-$3.3 (Vela), and SNR G266.2$-$1.2 (Vela~Jr.) are spatially overlapping SNRs. Among them, G263.9-3.3 is located in the foreground at a distance of 250–350 pc \citep{1999ApJ...515L..25C}. For Puppis~A, \citet{2003MNRAS.345..671R} suggested a distance of 2.2 kpc based on an H\,\textsc{I} depression near the NS RX~J0822$-$4300. 
    Later, \citet{2017MNRAS.464.3029R} analyzed H\,\textsc{I} absorption profiles and constrained the systemic velocity of Puppis~A to $+10.0\pm2.5$ \textrm{km~s$^{-1}$}, corresponding to a kinematic distance of $1.3\pm0.3$ kpc. In this work, we identified two extinction jumps at 0.877 kpc and 1.732 kpc. The far distance of 1.732 kpc falls within the previously reported range of 1.3-2.2 kpc. Considering the relative locations and analysis of Vela and Vela~Jr., we interpret the 1.732 kpc jump as the most probable distance to Puppis~A and classify it as Level A. The nearer jump at 0.877 kpc is likely associated with Vela Jr. 
    
    \item G263.9-3.3 (Vela). The most reliable distance estimate for this remnant probably comes from parallax measurements of its associated central Vela pulsar, which is located at $287\pm19$ pc \citep{2003ApJ...596.1137D}. However, no significant extinction jump is detected at this distance, likely due to the low dust content within the Local Bubble, the evolved and low-density nature of the SNR shell, and the absence of strong molecular cloud interactions. Instead, the two distances we identified (0.875 kpc and 1.788 kpc) are more likely associated with the nearby SNRs Puppis~A and Vela~Jr, respectively. Given the absence of an identifiable extinction jump corresponding to Vela itself, we are unable to constrain its distance using our method. This SNR is therefore classified as Level C. 
    
    \item G266.2$-$1.2 (Vela~Jr.) is a shell-type SNR that lies along the same line of sight as the Vela SNR, and appears fainter than Vela SNR in radio frequencies. The distance to G266.2$-$1.2 was determined to be $0.5-1.0$ kpc by \citet{2015ApJ...798...82A} primarily based on its measured X-ray expansion rate and hydrodynamic modeling. Their analysis also incorporated additional constraints, including consistency with the nearer component of the Vela Molecular Ridge and the Vel~OB1 association (0.8 kpc), as well as matching absorption column density, leading to a most likely distance estimate of $0.7\pm0.2$ kpc. In our work, we identified two extinction jumps at 0.843 kpc and 2.002 kpc. We infer that the closer distance of 0.843 kpc corresponds to G266.2$-$1.2, as it is consistent with the \citet{2015ApJ...798...82A}, and classify it as Level A, similar to Puppis~A. 

    \item G284.3$-$1.8 is a shell-type SNR and is potentially associated with the $\gamma$-ray binary 1FGL~J1018.6–5856 with a distance of $5.4^{+4.6}_{-2.1}$ kpc \citep{2011PASP..123.1262N}. Later, \citet{2019RAA....19...92S} used RCs to constrain the distance to $5.5\pm0.7$ kpc. However, \citet{1986ApJ...309..667R} used \(^{13}\text{CO}\) observations to identify interactions between the SNR and nearby MC and derived a kinematic distance of 2.9 kpc. We identified two extinction jumps at 0.398 kpc and 3.305 kpc. The large 3.305 kpc is consistent with \citet{1986ApJ...309..667R}. We therefore adopt 3.305 kpc as the distance to G284.3$-$1.8 and classify it as Level B. 

    \item G292.0+1.8 is a core-collapse SNR. Based on H\,\textsc{I} emission spectra and comparison of estimated hydrogen column density with X-ray absorption measurements, \citet{2003ApJ...594..326G} derived a kinematic distance of $6.2 \pm 0.9$~kpc. We identified two extinction jumps at 1.538~kpc and 4.073~kpc. The far distance at 4.073~kpc is likely related to the SNR. However, due to the lack of background stars beyond 6~kpc along this line of sight, we are unable to verify whether an additional extinction jump exists near $\sim$6.2~kpc. Consequently, we adopt 4.073 kpc as a lower-limit estimate and classify G292.0+1.8 as Level C due to this observational limitation.  
     
    \item G320.4$-$1.2 (RCW~89, MSH~15-52) is a complex SNR interacting with the young pulsar PSR~B1509–58. Based on H\,\textsc{I} absorption spectroscopy and supported by X-ray-derived column density gradients, \citet{1999MNRAS.305..724G} derived a distance of $5.2 \pm 1.4$~kpc for SNR G320.4$-$1.2. In this work, we obtained two extinction jumps at 0.998 kpc and 2.658 kpc. However, the lack of stars beyond 5 kpc in the \textit{Gaia} data limits to detect extinction jump at far distance. \cite{2020AA...639A..72W} used RCs as tracers and reported a distance of $3\pm0.04$, which is closer to our far extinction jump at 2.658 kpc. Furthermore, a significant number of stars in our sample show large color excess beyond 2 kpc, indicating substantial extinction in that distance range. We therefore adopt 2.658 kpc as a reasonable estimate for the distance to G320.4$-$1.2 and classify it to Level B. 
    
    \item G332.4-0.4 (RCW~103) is a shell type SNR. Based on 21 cm H\,\textsc{I} absorption measurements, \citet{1975A&A....45..239C} suggested a distance of 3.3 kpc. \citet{2004PASA...21...82R} confirmed this distance by analyzing H\,\textsc{I} line observations and placed the SNR in the nearby Scutum-Crux Arm. This distance is also consistent with the result of $3\pm0.3$ kpc obtained by \citet{2019RAA....19...92S} using RCs. We identified two extinction jumps at 1.054 and 2.504 kpc. Considering distance uncertainties, the difference between the larger value and previous estimates is within 25$\%$. Therefore, we adopt 2.504 kpc as the distance to RCW~103. Due to the clear extinction jump and good agreement with literature values, this SNR is classified as Level A. Given that RCW~103 and Kes~32 lie along the line of sight to the Norma Arm, the nearer extinction jump at 1.054 kpc may be attributed to dusty clouds in the foreground Scutum–Centaurus Arm. 
    
    \item G332.4+0.1 (Kes~32, MSH~16-51) lies along the line of sight to the Norma Arm, near G332.4-0.4 (RCW 103). \citet{2004ApJ...604..693V} reported that Kes 32 exhibits a higher hydrogen column density than RCW~103, suggesting that Kes~32 is located at a greater distance than RCW~103. While RCW~103 is located in the nearby Scutum–Crux Arm, Kes~32 was estimated to lie farther away, at a distance of 7.5–11 kpc in the Norma Arm, which is agreement with OH absorption measurements indicating a lower limit of 6.6 kpc \citep{1975MNRAS.173..649C}. We identified two extinction jumps at 1.167 and 2.972 kpc, both significantly closer than the previous distance estimates. Due to the lack of stars beyond 4 kpc, we are unable to assess the presence of a more distant extinction jump in the 7.5-11 kpc range. The two distances we detect are likely associated with Scutum-Crux Arm and Scutum-Centaurus Arm. Based on these considerations, Kes~32 is classified as Level C. 
    
    \item G338.3-0.0 is a composite SNR with a broken radio shell approximately 8$\arcmin$ in diameter. \citet{2016AA...589A..51S} inferred a distance in the range of 8.5-13 kpc. In contrast, we identified two extinction jumps at 0.928 kpc and 2.802 kpc, both significantly closer than the distances reported by \citet{2016AA...589A..51S}. Similar to the case of G332.4+0.1, these two distances may caused by foreground clouds in the Scutum-Crux Arm and Scutum-Centaurus Arm. Therefore, G338.3$-$0.0 is classified as Level C. 
    
    \item G359.1-0.5. We identified two extinction jumps at 1.106 kpc and 2.118 kpc. \citet{2020ApJ...893..147S} suggested a kinematic distance of 4 kpc, while \citet{2020AA...639A..72W} derived a distance of $3.29\pm0.47$ kpc using RCs. our sample lacks stars beyond 3 kpc along this line of sight, limiting our ability to probe extinction jump at greater distances. We adopt 2.118 kpc as the distance to G359.1$-$0.5, which falls within the 3$\sigma$ uncertainty of \citet{2020AA...639A..72W}, and classify this SNR as Level B. 
\end{enumerate}

To summarize, we analyzed 22 SNRs exhibiting two or more significant extinction jumps along their lines of sight. In several cases (e.g., G89.0+4.7, G160.9+2.9, and G260.4$-$3.4), our extinction--distance profiles not only agree well with reported kinematic results but also offer improved precision. This demonstrates that the extinction-based method, when combined with kinematic estimates, provides an independent and effective approach to verify and refine SNR distance determinations, even in complex Galactic environments. Moreover, our analysis yields significantly improved precision for many small-angular-size SNRs, highlighting the advantage of using \textit{Gaia}-based extinction profiles in distance estimates. In contrast, SNRs located in the Galactic first and fourth quadrants (e.g., G53.4+0.3, G54.+0.3, G338.3$-$0.0, and G332.4+0.1) lack distant stellar samples due to \textit{Gaia} DR3 observation limits. This restricts our ability to identify extinction jumps beyond $\sim$5 kpc, possibly leading to underestimated distances and reduced classification confidence. Overall, our analysis highlights both the strengths and limitations of extinction-based distance measurements, particularly for sightlines intersecting multiple spiral arms or MC complexes. Future studies combining deep infrared data and improved stellar parameter catalogs will be crucial for further refining distances to these challenging SNRs.

\subsection{Spatial Distribution of SNRs in the Galactic Plane}\label{sec:Spatial Distribution}
\begin{figure}[h]
\centering
\includegraphics[scale=0.8]{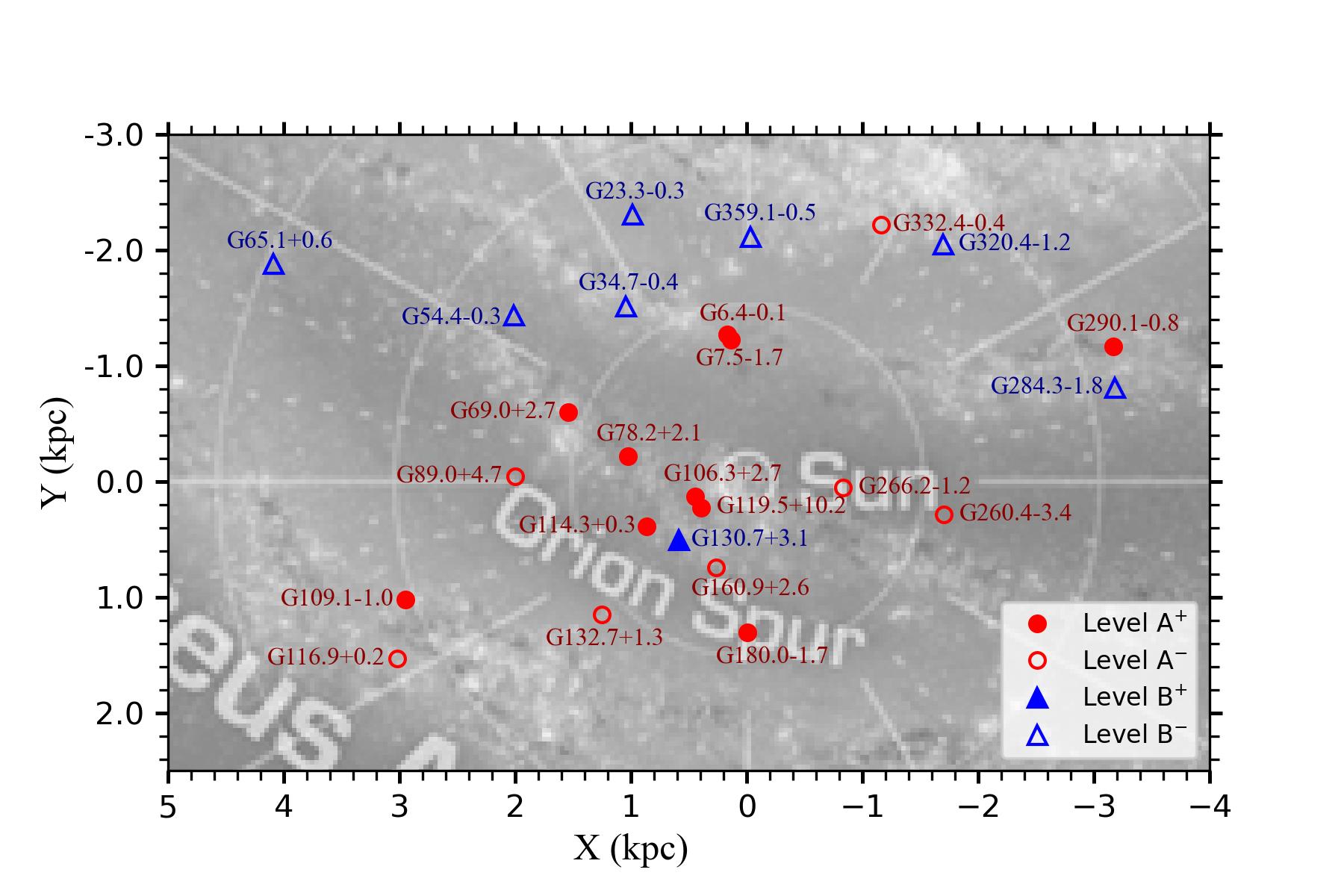}
\figurenum{3}
\caption{Spatial distribution of 17 Level A SNRs (red circles) and 8 Level B SNRs (blue triangles). Filled symbols (A$^+$/B$^+$) and open symbols (A$^-$/B$^-$) correspond to SNRs exhibiting a single clear extinction jump and two extinction jumps, respectively. The background image is adapted from \citet{2009PASP..121..213C}, centered on the Sun, with the Galactic Center oriented toward the top of the figure.}
\label{fig:4}
\end{figure}

Figure~\ref{fig:4} shows the Galactic plane projection for the 25 SNRs classified into Levels A and B, based on our derived distances. Due to observational limitations inherent to \textit{Gaia} DR3, stellar samples used for distance estimation are predominantly constrained to within $\sim$5 kpc. Consequently, our method is unable to accurately determine the distances for SNRs located beyond this range. Spatial analysis reveals that the majority of these well-constrained SNRs are located within the Local Arm, while the remaining two are located in the Perseus Arm.

Among the 17 SNRs classified as Level A, 12 (approximately 71\%) are located in or near the second and third Galactic quadrants. In these regions, extinction is generally lower and the line-of-sight distribution is relatively simple, which facilitates a clearer identification of extinction jumps. As a result, the distances we derived show excellent agreement with kinematic estimates from the literature. 
Our method further improves distance precision by leveraging direct \textit{Gaia} parallaxes and a sufficient number of stars along each sightline, enabling robust extinction--distance profiles. For example, in the case of SNR G89.0+4.7, we fill the gap of missing stars within 1.5--2.5 kpc noted in \citet{2020ApJ...891..137Z}, which had led to large uncertainty of distance to SNR. 
This consistency highlights the advantages of our extinction–distance method in combination with kinematic distances, particularly in areas of the Galaxy where the extinction environment is simple.

In contrast, SNRs located in the inner Galaxy are affected by kinematic distance ambiguity, i.e., each radial velocity along a given line of sight corresponds to two distances equally spaced on either side of the tangent point. This ambiguity poses a significant challenge to distance determination when relying on kinematic methods alone.
In our work, several Level C SNRs located in the inner Galaxy were previously reported to lie at the far kinematic distance. However, our extinction–distance analysis yields results closer to the near-side kinematic distance. For example, SNR G338.3$-$0.0 presents two prominent Local Standard of Rest (LSR) velocity components at $V_{\rm LSR} = -31$ and $-116$ \textrm{km~s$^{-1}$}, corresponding to near/far distances of 2.6/13.2 kpc and 6.5/9.3 kpc, respectively \citep{2016AA...589A..51S}. Our extinction-based estimate of 2.8 kpc is in close agreement with the near-side kinematic distance of 2.6 kpc associated with $V_{\rm{LSR}}=-31 \textrm{km~s$^{-1}$}$. This suggests that our extinction-based distance may correspond either to the genuine location of the SNR itself or to a foreground cloud along the sightline. Disentangling these possibilities requires further investigation. Extinction-based distances, when combined with kinematic estimates, provide complementary and independent constraints. These findings highlight that in the complex sightlines toward the inner Galactic disk, accurately determining SNR distances necessitates a complete characterization of the extinction profile, including both optical and infrared data, to robustly identify extinction jumps.

\subsection{Effect of Distance Binning on Derived Distances}\label{sec:binning_effect} 

\begin{figure}[ht]
\gridline{
  \fig{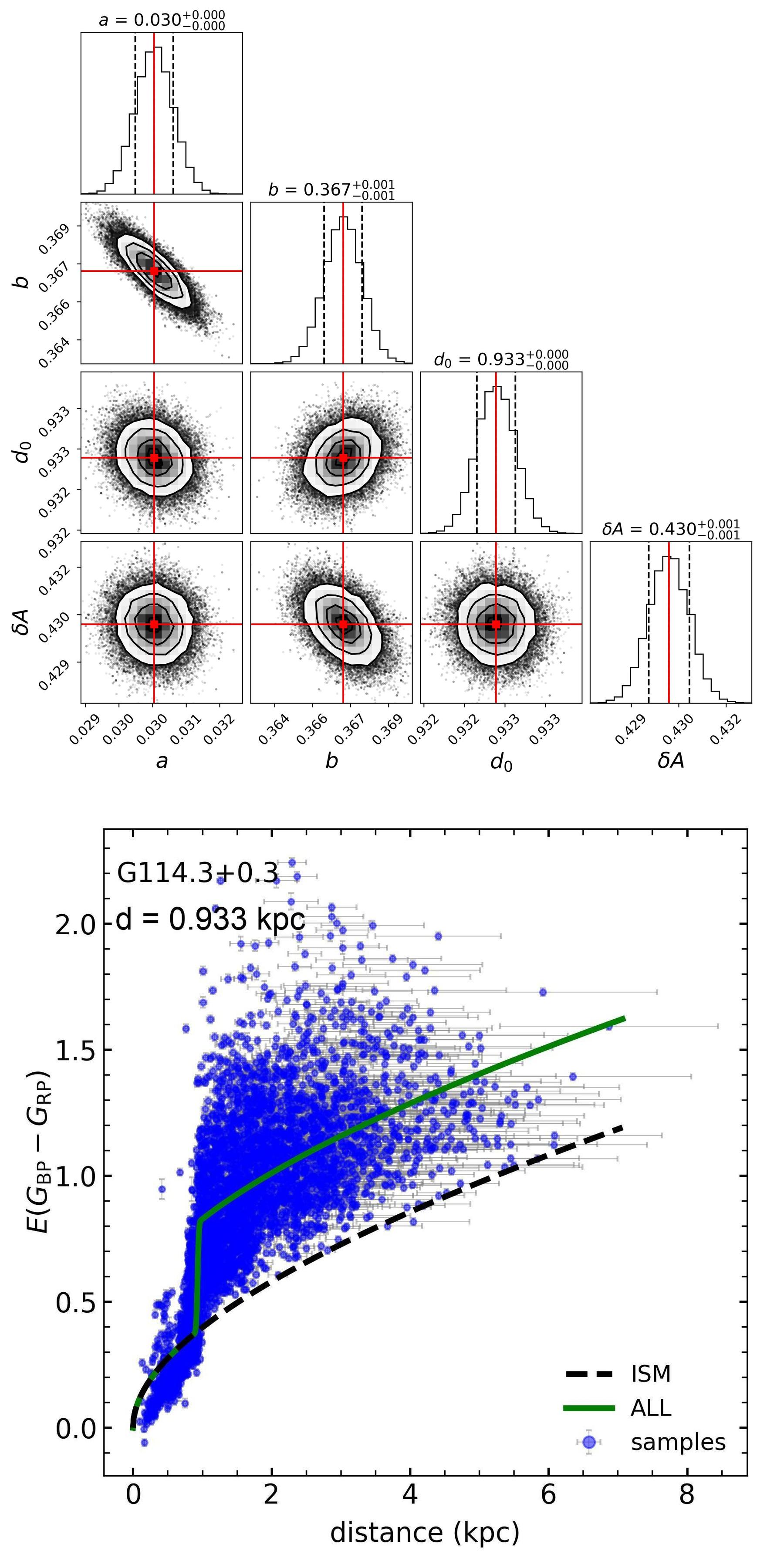}{0.42\textwidth}{(a) Unbinned}\hspace{-15mm}
  \fig{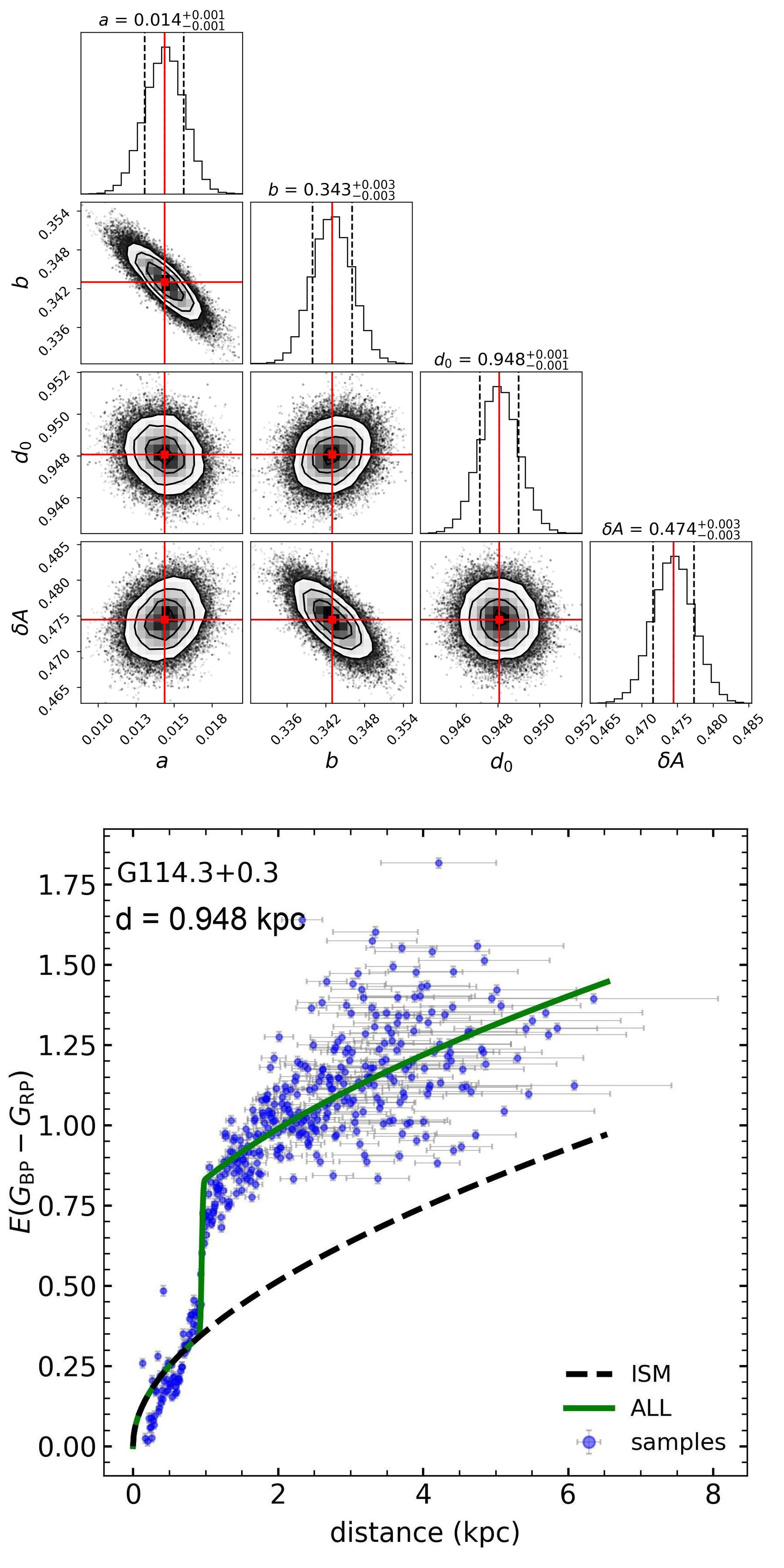}{0.42\textwidth}{(b) Binned}
}
\figurenum{4}
\caption{
Comparative analyses of G114.3+0.3 based on unbinned (left panels) and distance-binned (right panels) methodologies. The top panels display MCMC-derived corner plots characterizing the posterior distributions of the extinction--distance model parameters. The bottom panels show extinction versus distance profiles, where blue points represent individual stars within the SNR field, with gray error bars indicating uncertainties in both distance and color excess. The solid green lines are the best-fitting extinction profiles including both ISM and SNR contributions, while the black dashed lines denote the ISM-only component of the extinction model.
}
\label{fig:2}
\end{figure}

For SNRs with angular diameters larger than 20$\arcmin$, the number of stellar samples along their lines of sight is exceedingly high. To manage these densely populated stellar fields, we adopted a binning strategy in distance space, grouping stars into bins of 10 pc width. We confirmed that this binning approach does not introduce large systematic biases into the determination of SNR distances. For example, in the case of SNR G114.3+0.3, the derived distances from unbinned and binned analyses are 0.933 kpc and 0.948 kpc, respectively, differing by only about 2\%. A comparison between the unbinned case (Figure~\ref{fig:2}a) and the binned result (Figure~\ref{fig:2}b) shows that the binning method enhances the prominence of extinction jumps by improving the precision of both distance and extinction measurements.

\subsection{Effect of Different ISM Models on Derived Distances}\label{sec:different ISM}

\begin{figure*}[h]
\gridline{
  \fig{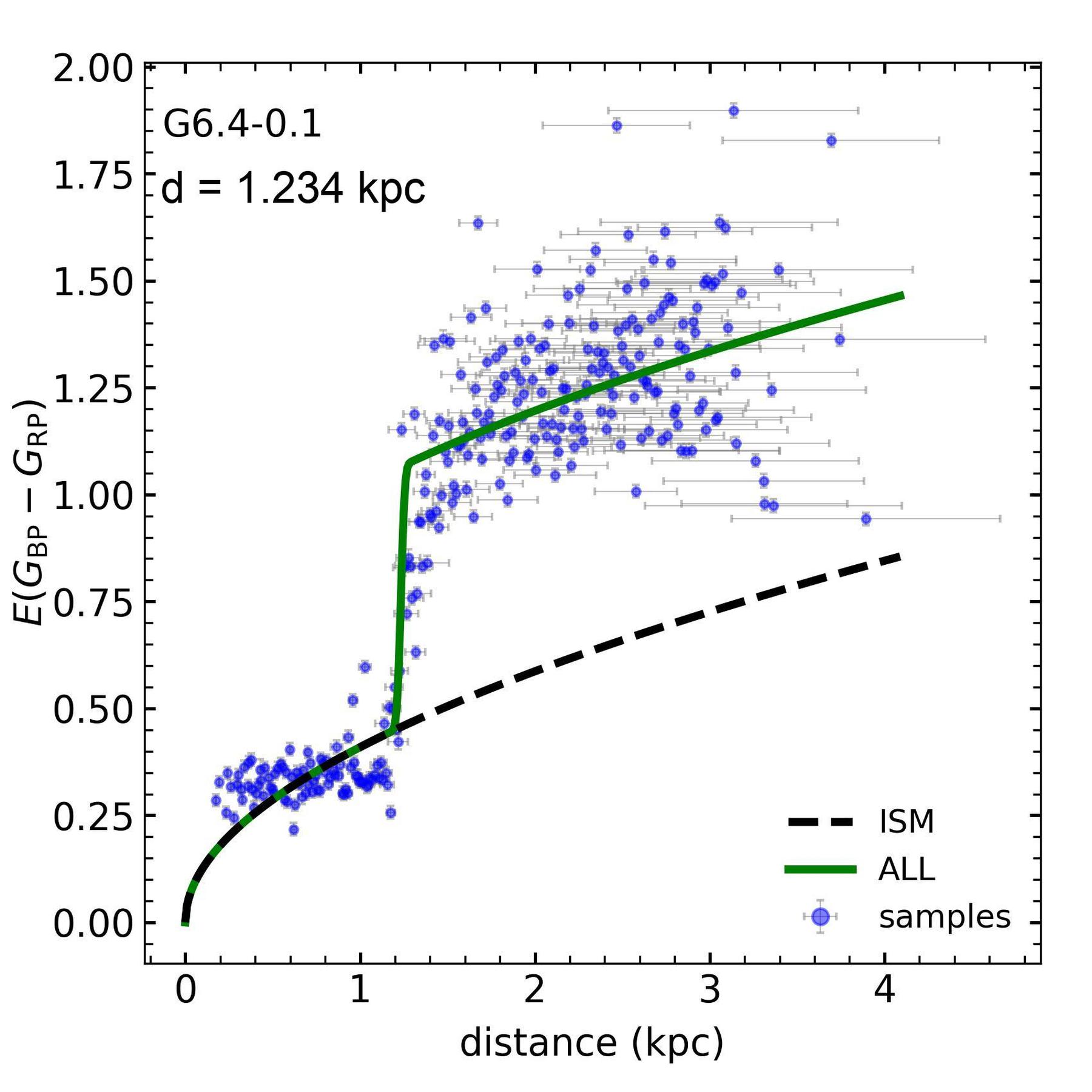}{0.45\textwidth}{(a) ISM in extinction--distance model}\hspace{-12mm}
  \fig{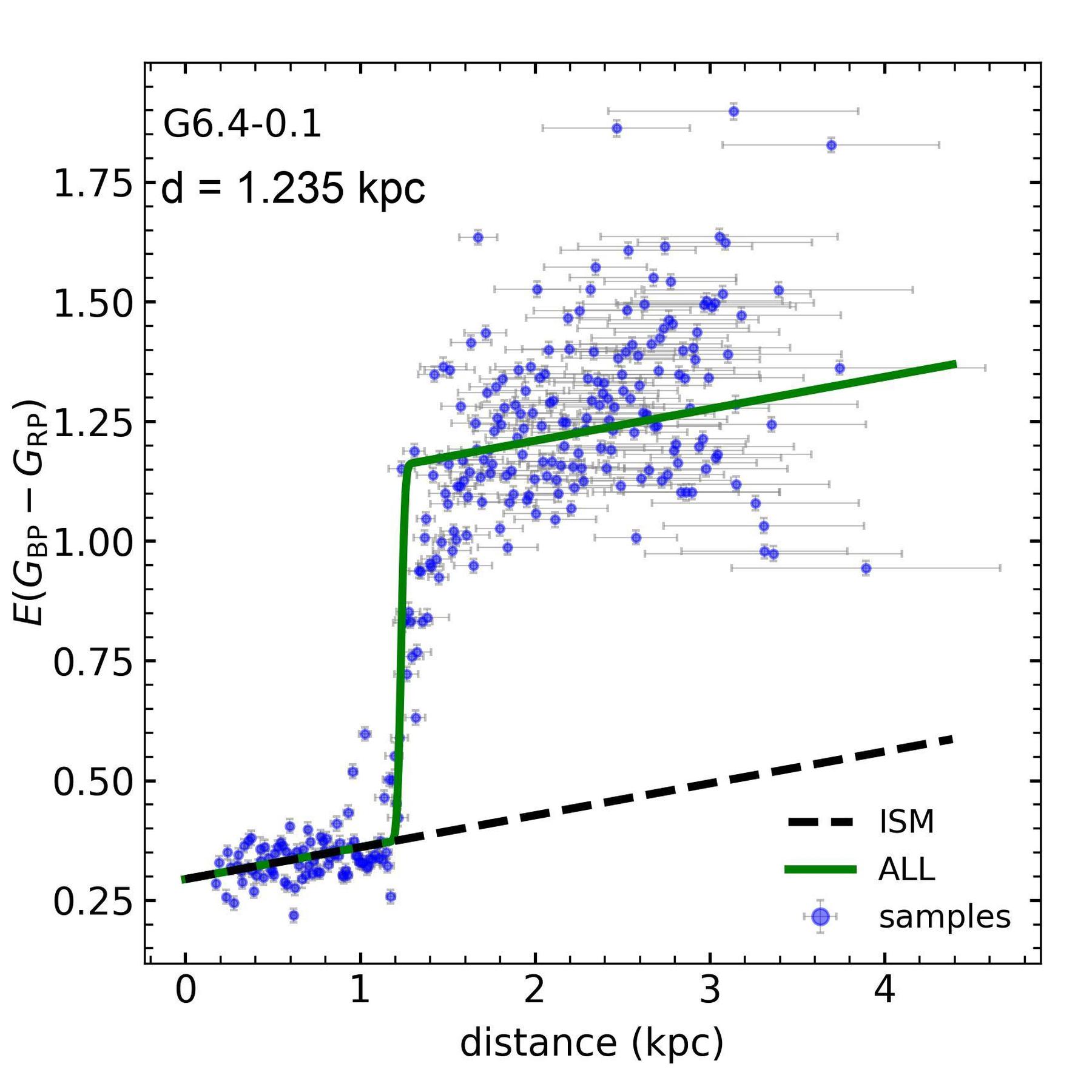}{0.45\textwidth}{(b) Independent linear foreground}
}
\figurenum{5}
\caption{
Color excess $E(G_{\rm BP}-G_{\rm RP})$ versus distance for SNR G6.4$-$0.1, modeled using the full extinction--distance method (left panel) and an independent linear foreground approximation (right panel). Blue points represent individual stars within the SNR field, with gray error bars indicating uncertainties in both distance and color excess. The solid green lines show the best-fitting extinction profiles including both diffuse ISM and SNR-associated components, while the black dashed lines represent the best-fit ISM-only contribution.
}
\label{fig:3}
\end{figure*}

Seventeen SNRs within $\sim$1 kpc exhibit nearly constant color excess values, with $E(G_{\rm BP}-G_{\rm RP})$ around 0.2--0.5 mag, indicating substantial foreground extinction. As described in Section~\ref{sec:method}, we addressed this by independently constructing a simplified linear foreground extinction model to replace the standard diffuse extinction term $A_{0}(d)$ in the MCMC fitting procedure. This modified approach retains sensitivity to detecting extinction jumps while providing a better matching to the observed extinction--distance profile. An illustrative example is SNR G6.4$-$0.1 (Figure~\ref{fig:3}), where both methods yield consistent distance estimates of 1.235 and 1.234 kpc, but the linear foreground extinction model clearly provides a superior fit to the extinction--distance profile. 

Considering the structure of the Milky Way, we also attempted to introduce an empirical model for the extinction caused by diffuse ISM, expressed as:
\begin{equation}
A_0(d) = a \cdot d - b \cdot \exp(-d/c) + k,
\end{equation}
This model incorporates an exponential decay term to describe the distribution of dust in the Galactic plane: extinction rises steeply at near distances and transitions into an approximately linear increase at far distances.
However, fits based on this model showed noticeable discrepancies with observed extinction--distance profiles for both ISM and SNRs. We therefore adopted the simpler empirical model given in Equation~\ref{equ1} of Section~\ref{sec:method}, which yields better agreement with data across multiple sightlines.

\section{summary}\label{sec:summary}

In this paper, we developed a systematic extinction--distance method to determine robust distances for 44 NS-associated SNRs.
Leveraging high-quality \textit{Gaia} DR3 photometric and astrometric data, combined with reliable stellar parameters from the SHBoost catalog, we constructed extinction--distance profiles along each SNR line of sight and applied a combined two-components extinction model (diffuse ISM and SNR-associated extinction) with MCMC fitting to identify the most probable distances. 

We obtained reliable extinction--distance profiles for all 44 SNRs and classified the results into three reliability levels: 17 Level A SNRs, showing significant extinction jumps and excellent agreement with robust kinematic distances; 8 Level B SNRs, well consistent with results from other independent methods (e.g., RCs, X-ray absorption, $\Sigma-\rm{D}$); and 19 Level C SNRs, for which distance constraints are limited by insufficient background stars beyond 5 kpc or by the unclear extinction jumps. 
We validated the stability of our method against potential biases from distance binning and foreground ISM modeling. For SNRs with complex or foreground-dominated extinction features, adopting an independent linear foreground model improves the fit.

Compared to other SNR distance determination methods, our approach offers several key advantages:  
(1) It can effectively constrain distances to small-angular-size SNRs (down to $\sim$10$\arcmin$), which are inaccessible to RC–based methods due to limited stellar sampling. 
(2) By utilizing Gaia parallaxes, our method directly obtained stellar distances, enabling higher-precision extinction–-distance profiles than traditional photometric or statistical distances, especially effective in the second and third Galactic quadrants. 
(3) Our extinction-based approach provides an independent and complementary constraints, significantly enhancing the reliability of SNR distance determinations, particularly where kinematic distance ambiguity exists. 

Although our method currently performs less effectively in the inner Galaxy (first and fourth quadrants) due to high extinction, complex ISM structures, and the limited depth of Gaia observations, it is expected to improve with future releases of deeper parallax catalogs (e.g., Gaia DR4) and complementary infrared surveys. These forthcoming datasets will enhance the ability to detect distant extinction jumps and enable more complete extinction--distance profiles, thereby facilitating more accurate mapping of Galactic SNRs and their associated NSs.

\begin{acknowledgments}
We thank the anonymous referee for very useful comments/suggestions. We thank Dr. He Zhao, Prof. Biwei Jiang, and Tao Wang for very helpful discussion. This work is supported by the National Natural Science Foundation of China (NSFC) through the projects 12373028, 12322306, 12173047, and 12133002. This work is also supported by the science research grants from the China Manned Space Project with No. CMS-CSST-2025-A01. S.W. and X.C. acknowledge support from the Youth Innovation Promotion Association of the CAS (grant Nos. 2023065 and 2022055). This work has made use of data from the European Space Agency (ESA) mission Gaia (\url{https://www.cosmos.esa.int/gaia}),
processed by the Gaia Data Processing and Analysis Consortium (DPAC, \url{https://www.cosmos.esa.int/web/gaia/dpac/consortium}).
Funding for the DPAC has been provided by national institutions, in particular the institutions participating in the Gaia Multilateral Agreement.

\end{acknowledgments}

\begin{deluxetable*}{lccccccccccc}
\tablewidth{0pt}
\tablecaption{Distances of SNRs \label{tab:table1}}
\tablehead{
\colhead{SNR name} & \colhead{RA}    & \colhead{DEC}   & \colhead{D}     & \colhead{$N^{a}$}   &  \colhead{d$_{\rm this work}$} & \colhead{$\delta A$} & \colhead{d$_{\rm literature}$} & \colhead{Method} & \colhead{Ref} & \colhead{level}\\
\colhead{}         & \colhead{(deg)} & \colhead{(deg)} & \colhead{(arcminute)} & \colhead{}  & \colhead{(kpc)}               & \colhead{(mag)}      & \colhead{(kpc)}                & \colhead{}       & \colhead{}    & \colhead{}
}
\startdata
$\rm G0.9+0.1$ & 266.84 & -28.15 & 8 & 23 & $1.441\pm0.128$ & $0.639\pm0.006$ & 8.5, 13 & $\rm GC^{b}$, associated object & 1, 2 & $\rm {C^{\star}}^{e}$ \\
$\rm G6.4-0.1$ & 270.13 & -23.43 & 48 & 1348 & $1.235\pm0.001$ & $0.782\pm0.001$ & $1.9\pm0.3$, $3.55\pm0.9$ & kinematic, $\rm RCs^{c}$ & 3, 4 & $\rm {A^{+}}$ \\
$\rm G7.5-1.7$ & 272.5 & -23.17 & 100 & 6940 & $1.284\pm0.001$ & $0.661\pm0.003$ & 1.7 & associated object & 5 & $\rm {A^{+}}$ \\
$\rm G8.7-0.1$ & 271.38 & -21.43 & 45 & 411 & $1.382\pm0.001$ & $1.110\pm0.001$ & $4.15\pm0.19$, 4.5 & RCs, kinematic & 4, 6 & $\rm C^{\star}$ \\
$\rm G23.3-0.3$ & 278.69 & -8.8  & 27 & 428 & $2.508\pm0.002$ & $0.402\pm0.001$ & $3.38\pm0.26$, $4.8\pm0.2$, 2.7 &  RCs, kinematic, $\Sigma-\rm{D}$ & 4, 7, 8 & $\rm {B^{-}}$  \\
$\rm G32.8-0.1$ & 282.85 & -0.13 & 15 & 104 & $0.569\pm0.053$ & $1.235\pm0.021$ & $4.8\pm0.3$ & kinematic & 7 & $\rm C^{\star}$ \\
$\rm G33.6+0.1$ & 283.2 & 0.68 & 10 & 38 & $1.995\pm0.002$ & $0.801\pm0.001$ & $3.5\pm0.3$, 7.1 & kinematic, $\rm MC^{d}$ & 7, 9 & C \\
$\rm G34.7-0.4$ & 284.0 & 1.27 & 27 & 235 & $1.843\pm0.246$ & $0.402\pm0.239$ & $2.66\pm0.71$, $3\pm0.3$ &  RCs, kinematic & 4, 7 & $\rm {B^{-}}$ \\
$\rm G35.6-0.4$ & 284.48 & 2.22  & 11 & 21 & $1.590\pm0.001$ & $1.701\pm0.001$ & $3.8\pm0.3$, $3.6\pm0.4$ & kinematic, MC & 7, 10 & C \\
$\rm G53.4+0.0$ & 292.49 & 18.17 & 10 & 37 & $1.245\pm0.075$ & $0.588\pm0.163$ & 7.5 & kinematic & 11 & $\rm C^{\star}$ \\
$\rm G54.1+0.3$ & 292.63 & 18.87 & 12 & 88 & $1.405\pm0.004$ & $0.583\pm0.007$ &$ 4.9\pm0.8$, 5.6--7.2 & kinematic & 7, 12 & $\rm C^{\star}$ \\
$\rm G54.4-0.3$ & 293.33 & 18.93 & 40 & 1415 & $2.477\pm0.003$ & $0.290\pm0.004$ & $6.64\pm1.25$, 2.5, $6.6\pm0.6$ & RCs, $\Sigma-\rm{D}$, kinematic & 4, 8, 13 & $\rm {B^{-}}$ \\
$\rm G57.2+0.8$ & 293.75 & 21.95 & 12 & 54 & $0.650\pm0.048$ & $1.271\pm0.014$ & $6.6\pm0.7$, 6.6 & kinematic, MC & 14, 15 & $\rm C^{\star}$ \\
$\rm G65.1+0.6$ & 298.67 & 28.58 & 50 & 3263 & $4.500\pm0.498$ & $0.410\pm0.001$ & $4.16\pm0.61$, 2.6, $9.3\pm0.3$ & RCs, $\Sigma-\rm{D}$, kinematic & 4, 8, 16 & $\rm {B^{-}}$ \\
$\rm G69.0+2.7$ & 298.33 & 32.92 & 80 & 8968 & $1.654\pm0.003$ & $0.385\pm0.002$ & 1.5, $4.6\pm0.8$ & kinematic, RCs & 17, 18 & $\rm {A^{+}}$ \\
$\rm G78.2+2.1$ & 305.21 & 40.43 & 60 & 1543 & $1.051\pm0.001$ & $1.236\pm0.001$ & 1.2, 0.98, $2.1\pm0.4$ & $\Sigma-\rm{D}$, extinction, kinematic & 8, 19, 20 & $\rm {A^{+}}$ \\
$\rm G89.0+4.7$ & 311.25 & 50.58 & 90 & 4409 & $2.010\pm0.003$ & $0.420\pm0.002$ & $1.6\pm0.2$, $1.9^{+0.3}_{-0.2}$, $2.3\pm0.3$ & kinematic, RCs, extinction & 15, 18, 19 & $\rm {A^{-}}$ \\
$\rm G106.3+2.7$ & 336.88 & 60.83 & 24 & 300 & $0.465\pm0.001$ & $0.699\pm0.004$ & $0.8\pm0.1$, 0.8 & MC, kinematic & 15, 21 & $\rm {A^{+}}$ \\
$\rm G109.1-1.0$ & 345.40 & 58.88 & 28 & 561 & $3.122\pm0.003$ & $0.441\pm0.005$ & $2.79\pm0.04$, $3.1\pm0.2$ & extinction, kinematic & 19, 22 & $\rm {A^{+}}$ \\
$\rm G114.3+0.3$ & 354.25 & 61.92 & 55 & 1812 & $0.948\pm0.001$ & $0.474\pm0.003$ & 0.7, 3.4 & kinematic & 23, 24 & $\rm {A^{+}}$ \\
$\rm G116.9+0.2$ & 359.79 & 62.43 & 34 & 634 & $3.383\pm0.006$ & $0.087\pm0.012$ & 1.6, 3.383 & kinematic & 23, 25 & $\rm {A^{-}}$ \\
$\rm G119.5+10.2$ & 1.67 & 72.75 & 90 & 2999 & $0.462\pm0.004$ & $0.420\pm0.003$ & 1.5, $1.4\pm0.3$ & $\Sigma-\rm{D}$, kinematic & 8, 26 & $\rm {A^{+}}$ \\
$\rm G130.7+3.1$ & 31.42 & 64.82 & 9 & 17 & $0.774\pm0.255$ & $0.632\pm0.029$ & $2\pm0.3$ & Perseus Arm & 27 & $\rm {B^{+}}$ \\
$\rm G132.7+1.3$ & 34.42 & 62.75 & 80 & 3881 & $1.701\pm0.001$ & $0.210\pm0.001$ & $1.95\pm0.04$ & kinematic & 28 & $\rm {A^{-}}$ \\
$\rm G160.9+2.6$ & 75.25 & 46.67 & 120 & 7268 & $0.791\pm0.004$ & $0.192\pm0.003$ & $0.54\pm0.1$, $0.8\pm0.4$ & extinction, kinematic & 19, 29 & $\rm {A^{-}}$ \\
$\rm G180.0-1.7$ & 84.75 & 27.83 & 180 & 14093 & $1.303\pm0.006$ & $0.203\pm0.003$ & $1.33^{+0.103}_{-0.112}$, 1.37 & kinematic & 30, 31 & $\rm {A^{+}}$ \\
$\rm G260.4-3.4$ & 125.5 &-43 & 50 & 1250 & $1.732\pm0.001$ & $0.410\pm0.003$ & $1.3\pm0.3$, 2.2 & kinematic & 32, 33 & $\rm {A^{-}}$ \\
$\rm G263.9-3.3$ & 128.5 & -45.83 & 255 & 28564 & -- & -- & $0.25\pm0.03$, $0.29\pm0.02$ & parallax, associated object & 34, 35 & C \\
$\rm G266.2-1.2$ & 133  & -46.33 & 120 & 8749 & $0.836\pm0.002$ & $0.630\pm0.003$ & $0.7\pm0.2$ & MC & 36 & $\rm {A^{-}}$ \\
$\rm G284.3-1.8$ & 154.56 & -59 & 24 & 620 & $3.287\pm0.004$ & $0.732\pm0.003$ & $5.5\pm0.7$, 2.9, $5.6^{+4.6}_{-2.1}$ & RCs, kinematic, associated objects & 37, 38, 39 & $\rm {B^{-}}$ \\
\enddata
\textbf{Notes:} \\
$\rm ^a$ N: Number of stars along the SNR sightline used in constructing the extinction–distance profile. $\rm ^b$ GC: Galactic center. $\rm ^c$ RCs: Red clump stars. $\rm ^d$ MC: molecular cloud association/interaction. $\rm ^e$ $\rm {C^{\star}}$: SNRs in Level C that lack distant stellar samples. \\
\end{deluxetable*}

\addtocounter{table}{-1}
\begin{deluxetable*}{lcccccccccc}
\tablecaption{(continued) \label{tab:table1}}
\tablewidth{0pt}
\tablehead{
\colhead{SNR name} & \colhead{RA}    & \colhead{DEC}   & \colhead{D}     & \colhead{$N^{a}$}   &  \colhead{d$_{\rm this work}$} & \colhead{$\delta A$} & \colhead{d$_{\rm literature}$} & \colhead{Method} & \colhead{Ref} & \colhead{level}\\
\colhead{}         & \colhead{(deg)} & \colhead{(deg)} & \colhead{(arcminute)} & \colhead{}  & \colhead{(kpc)}               & \colhead{(mag)}      & \colhead{(kpc)}                & \colhead{}       & \colhead{}    & \colhead{}
}
\startdata
$\rm G290.1-0.8$ & 165.77 & -60.93 & 14 & 262 & $3.377\pm0.002$ & $0.268\pm0.002$ & $7\pm1$, $6.3\pm0.8$, 3.4--4 & kinematic & 40, 41, 42 & $\rm {A^{+}}$ \\
$\rm G292.0+1.8$ & 171.15 & -59.27 & 8 & 74 & $4.074\pm0.008$ & $0.196\pm0.007$ & $6.2\pm0.9$ & kinematic & 43 & $\rm C^{\star}$ \\
$\rm G292.2-0.5$ & 169.83 & -61.47 & 15 & 202 & $2.422\pm0.003$ & $0.869\pm0.002$ & $8.4\pm0.4$ & kinematic & 44 & $\rm C^{\star}$ \\
$\rm G296.5+10.0$ & 182.42 & -52.42 & 65 & 2520 & -- & -- & $3.8\pm0.5$, $2.1^{+1.8}_{-0.8}$ & RCs, kinematic & 4, 45 & C \\
$\rm G296.8-0.3$ & 179.6 & -62.6 & 14 & 211 & $3.294\pm0.122$ & $0.535\pm0.030$ & $9.6\pm0.6$ & kinematic & 46 & $\rm C^{\star}$ \\
$\rm G320.4-1.2$ & 228.6 & -59.1 & 35 & 1360 & $2.658\pm0.003$ & $0.501\pm0.001$ & $3\pm0.4$, $5.2\pm1.4$ &  RCs, kinematic & 4, 47 & $\rm {B^{-}}$ \\
$\rm G332.4+0.1$ & 243.8 & -50.7 & 15 & 259 & $2.972\pm0.002$ & $0.801\pm0.001$ & 7.5--11 & Norma Arm & 48 & $\rm C^{\star}$\\
$\rm G332.4-0.4$ & 244.4 & -51.0 & 10 & 99 & $2.504\pm0.133$ & $0.810\pm0.001$ & $3\pm0.3$, 3.3, 6.5 & extinction, kinematic, extinction & 37, 49, 50 & $\rm {A^{-}}$ \\
$\rm G338.3-0.0$ & 250.25 & -46.57 & 8 & 50 & $2.802\pm0.938$ & $1.316\pm0.484$ & 8.5--13 & kinematic & 51 & $\rm C^{\star}$ \\
$\rm G338.5+0.1$ & 250.29 & -46.32 & 9 & 64 & $1.062\pm0.040$ & $0.736\pm0.004$ & 11 & Norma Arm & 52 & $\rm C^{\star}$ \\
$\rm G344.7-0.1$ & 255.96 &-41.7 & 8 & 39 & $1.250\pm0.050$ & $0.676\pm0.013$ & $6.3\pm0.1$, 14 & kinematic, $\Sigma-\rm{D}$ & 53, 54 & $\rm C^{\star}$ \\
$\rm G348.7+0.3$ & 258.48 & -38.18 & 17 & 139 & $1.618\pm0.001$ & $1.130\pm0.003$ & $13.2\pm0.2$ & kinematic & 55 & $\rm C^{\star}$\\
$\rm G351.7+0.8$ & 260.25 & -35.45 & 14 & 74 & $1.094\pm0.001$ & $0.566\pm0.003$ & $3.35\pm0.11$, 5.4, $13.2\pm0.5$ & RCs, $\Sigma-\rm{D}$, kinematic & 4, 8, 56 & $\rm C^{\star}$\\
$\rm G359.1-0.5$ & 266.38 & -29.95 & 24 & 183 & $2.118\pm0.507$ & $0.464\pm0.125$ & $3.29\pm0.47$, 5  & RCs, kinematic & 4, 57 & $\rm {B^{-}}$ \\
\enddata
\textbf{Notes:} \\
$\rm ^a$ N: Number of stars along the SNR sightline used in constructing the extinction–distance profile. $\rm ^b$ GC: Galactic center. $\rm ^c$ RCs: Red clump stars. $\rm ^d$ MC: molecular cloud association/interaction. $\rm ^e$ $\rm {C^{\star}}$: SNRs in Level C that lack distant stellar samples. \\
\textbf{References:}
(1)\citet{2005AA...432L..25A}, (2)\citet{2009ApJ...700L..34C}, (3)\citet{2002AJ....124.2145V}, (4)\citet{2020AA...639A..72W}, (5)\citet{2008ApJ...681..320R}, (6)\citet{2009ApJ...694L..16H},
(7)\citet{2018AJ....155..204R}, 
(8)\citet{2013ApJS..204....4P}, 
(9)\citet{2016ApJ...816....1K}, 
(10)\citet{2013ApJ...775...95Z}, 
(11)\citet{2018ApJ...860..133D}, 
(12)\citet{2008AJ....136.1477L}, 
(13)\citet{2017ApJ...843..119R}, 
(14)\citet{2020ApJ...905...99Z}, 
(15)\citet{2023ApJS..268...61Z}, 
(16)\citet{2006AA...455.1053T}, 
(17)\citet{2012MNRAS.423..718L}, 
(18)\citet{2018ApJS..238...35S}, 
(19)\citet{2020ApJ...891..137Z}, 
(20)\citet{2013MNRAS.436..968L}, 
(21)\citet{2001ApJ...560..236K}, 
(22)\citet{2018MNRAS.473.1705S}, 
(23)\citet{2004ApJ...616..247Y}, 
(24)\citet{1986ApJ...303..465F}, 
(25)\citet{1981AA....99...17R}, 
(26)\citet{1993AJ....105.1060P}, 
(27)\citet{2013AA...560A..18K}, 
(28)\citet{2016ApJ...833....4Z}, 
(29)\citet{2007AA...461.1013L}, 
(30)\citet{2015MNRAS.448.3196D}, 
(31)\citet{2024ApJ...968...94K}, 
(32)\citet{2017MNRAS.464.3029R},
(33)\citet{2003MNRAS.345..671R}, 
(34)\citet{1999ApJ...515L..25C}, 
(35)\citet{2003ApJ...596.1137D}, 
(36)\citet{2015ApJ...798...82A}, 
(37)\citet{2019RAA....19...92S}, 
(38)\citet{1986ApJ...309..667R}, 
(39)\citet{2011PASP..123.1262N}, 
(40)\citet{2006MNRAS.369..416R}, 
(41)\citet{2022ApJ...940...63R},
(42)\citet{1973ApL....15...61D}, 
(43)\citet{2003ApJ...594..326G}, 
(44)\citet{2004MNRAS.352.1405C}, 
(45)\citet{2000AJ....119..281G}, 
(46)\citet{1998MNRAS.296..813G}, 
(47)\citet{1999MNRAS.305..724G},  
(48)\citet{2004ApJ...604..693V}, 
(49)\citet{2004PASA...21...82R}, 
(50)\citet{1983AJ.....88.1210R}, 
(51)\citet{2016AA...589A..51S}, 
(52)\citet{2007AA...468..993K}, 
(53)\citet{2011AA...531A.138G}, 
(54)\citet{1993AJ....105.2251D}, 
(55)\citet{2012MNRAS.421.2593T}, 
(56)\citet{2007MNRAS.378.1283T},
(57)\citet{2011MmSAI..82..703F}, 
\end{deluxetable*}

\begin{figure}[t]
\centering
\includegraphics[scale=0.196]{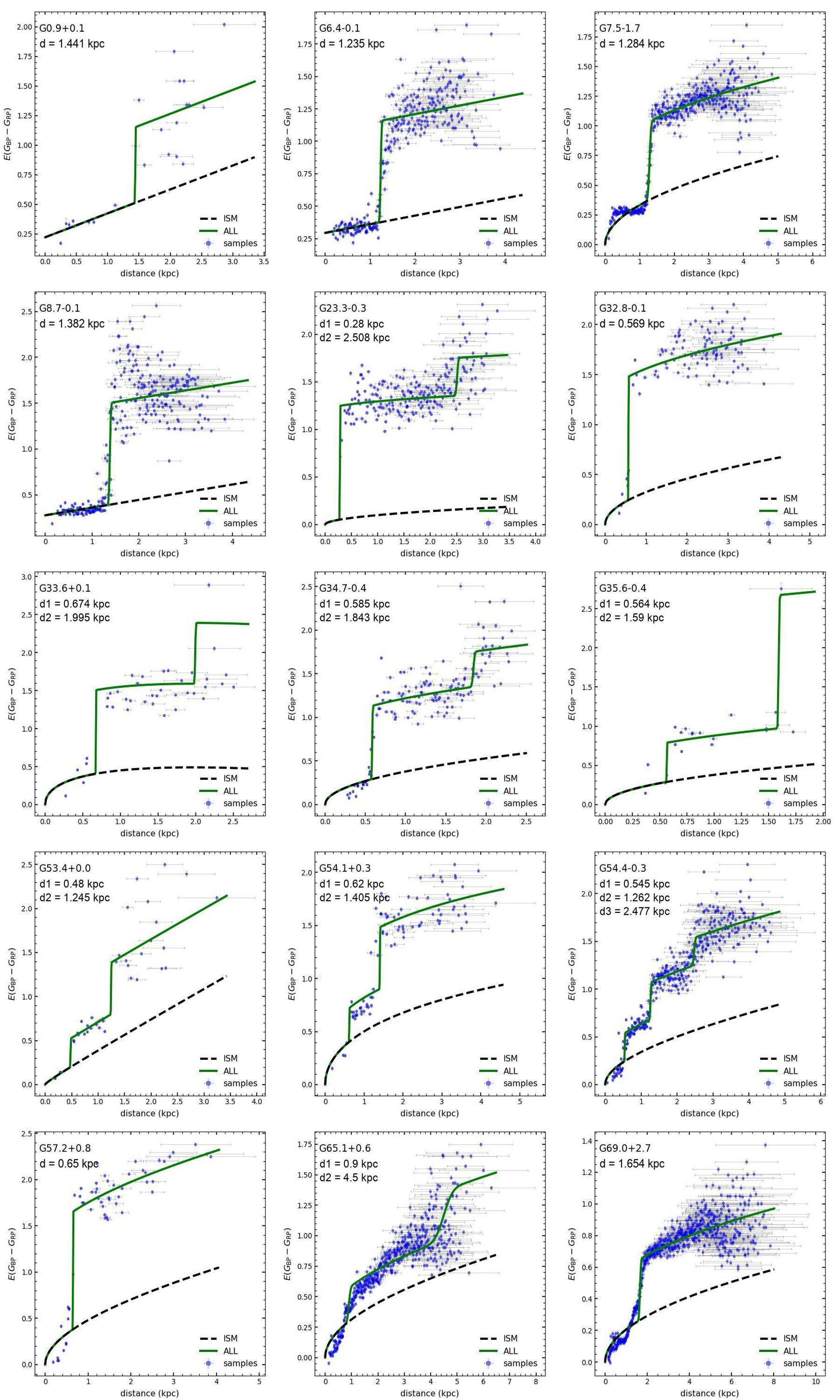}
\figurenum{1}
\caption{Color excess $E(G_{\rm BP}-G_{\rm RP})$ versus distance for all 44 SNRs. Blue points represent individual stars within each SNR field, with gray error bars indicating uncertainties in both $E(G_{\rm BP}-G_{\rm RP})$ and distance. The solid green lines are the best-fitting extinction profiles derived from the extinction--distance model, including both diffuse ISM and SNR-associated extinction components. For comparison, the black dashed lines represent the best-fit ISM-only extinction profiles, either directly modeled or approximated using an independent linear foreground model in certain cases.}
\label{fig:1}
\end{figure}

\begin{figure}
\centering
\includegraphics[scale=0.196]{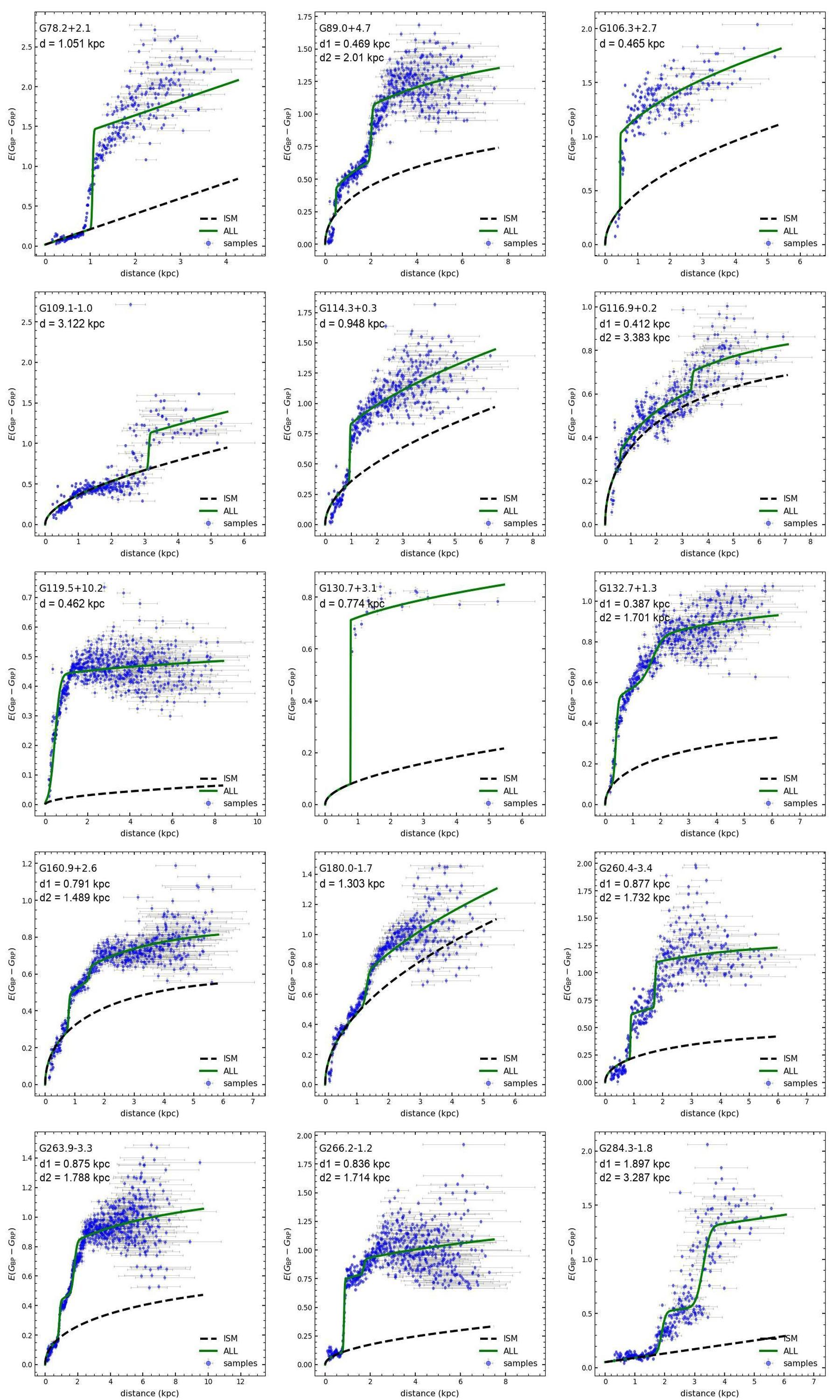}
\figurenum{1}
\caption{(continued).}
\end{figure}

\begin{figure}
\centering
\includegraphics[scale=0.196]{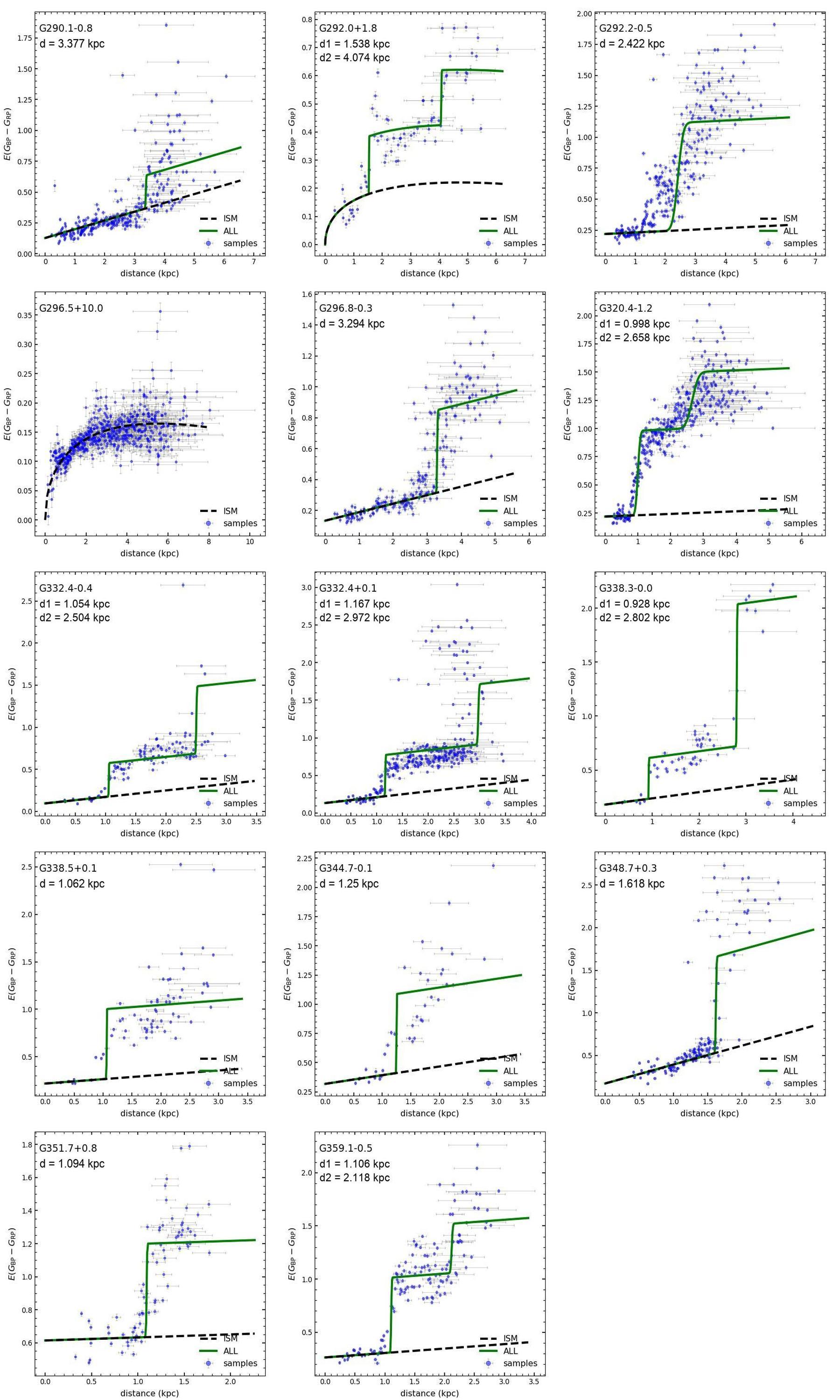}
\figurenum{1}
\caption{(continued).}
\end{figure}

\clearpage


\begin{thebibliography}{}
\expandafter\ifx\csname natexlab\endcsname\relax\def\natexlab#1{#1}\fi
\providecommand{\url}[1]{\href{#1}{#1}}
\providecommand{\dodoi}[1]{doi:~\href{http://doi.org/#1}{\nolinkurl{#1}}}
\providecommand{\doeprint}[1]{\href{http://ascl.net/#1}{\nolinkurl{http://ascl.net/#1}}}
\providecommand{\doarXiv}[1]{\href{https://arxiv.org/abs/#1}{\nolinkurl{https://arxiv.org/abs/#1}}}

\bibitem[{F. {Aharonian} {et~al.}(2005){Aharonian}, {Akhperjanian}, {Aye}, {Bazer-Bachi}, {Beilicke}, {Benbow}, {Berge}, {Berghaus}, {Bernl{\"o}hr}, {Boisson}, {Bolz}, {Borgmeier}, {Braun}, {Breitling}, {Brown}, {Bussons Gordo}, {Chadwick}, {Chounet}, {Cornils}, {Costamante}, {Degrange}, {Djannati-Ata{\"\i}}, {O'C. Drury}, {Dubus}, {Ergin}, {Espigat}, {Feinstein}, {Fleury}, {Fontaine}, {Funk}, {Gallant}, {Giebels}, {Gillessen}, {Goret}, {Hadjichristidis}, {Hauser}, {Heinzelmann}, {Henri}, {Hermann}, {Hinton}, {Hofmann}, {Holleran}, {Horns}, {de Jager}, {Jung}, {Kh{\'e}lifi}, {Komin}, {Konopelko}, {Latham}, {Le Gallou}, {Lemi{\`e}re}, {Lemoine}, {Leroy}, {Lohse}, {Marcowith}, {Masterson}, {McComb}, {de Naurois}, {Nolan}, {Noutsos}, {Orford}, {Osborne}, {Ouchrif}, {Panter}, {Pelletier}, {Pita}, {P{\"u}hlhofer}, {Punch}, {Raubenheimer}, {Raue}, {Raux}, {Rayner}, {Redondo}, {Reimer}, {Reimer}, {Ripken}, {Rob}, {Rolland}, {Rowell}, {Sahakian}, {Saug{\'e}}, {Schlenker}, {Schlickeiser}, {Schuster}, {Schwanke},
  {Siewert}, {Sol}, {Steenkamp}, {Stegmann}, {Tavernet}, {Terrier}, {Th{\'e}oret}, {Tluczykont}, {Vasileiadis}, {Venter}, {Vincent}, {Visser}, {V{\"o}lk}, \& {Wagner}}]{2005AA...432L..25A}
{Aharonian}, F., {Akhperjanian}, A.~G., {Aye}, K.~M., {et~al.} 2005, \bibinfo{title}{{Very high energy gamma rays from the composite SNR G 0.9+0.1},} \aap, 432, L25, \dodoi{10.1051/0004-6361:200500022}

\bibitem[{G.~E. {Allen} {et~al.}(2015){Allen}, {Chow}, {DeLaney}, {Filipovi{\'c}}, {Houck}, {Pannuti}, \& {Stage}}]{2015ApJ...798...82A}
{Allen}, G.~E., {Chow}, K., {DeLaney}, T., {et~al.} 2015, \bibinfo{title}{{On the Expansion Rate, Age, and Distance of the Supernova Remnant G266.2-1.2 (Vela Jr.)},} \apj, 798, 82, \dodoi{10.1088/0004-637X/798/2/82}

\bibitem[{R. {Andrae} {et~al.}(2023){Andrae}, {Rix}, \& {Chandra}}]{2023ApJS..267....8A}
{Andrae}, R., {Rix}, H.-W., \& {Chandra}, V. 2023, \bibinfo{title}{{Robust Data-driven Metallicities for 175 Million Stars from Gaia XP Spectra},} \apjs, 267, 8, \dodoi{10.3847/1538-4365/acd53e}

\bibitem[{M. {Andriantsaralaza} {et~al.}(2022){Andriantsaralaza}, {Ramstedt}, {Vlemmings}, \& {De Beck}}]{2022A&A...667A..74A}
{Andriantsaralaza}, M., {Ramstedt}, S., {Vlemmings}, W.~H.~T., \& {De Beck}, E. 2022, \bibinfo{title}{{Distance estimates for AGB stars from parallax measurements},} \aap, 667, A74, \dodoi{10.1051/0004-6361/202243670}

\bibitem[{W.~F. {Brisken} {et~al.}(2003){Brisken}, {Thorsett}, {Golden}, \& {Goss}}]{2003ApJ...593L..89B}
{Brisken}, W.~F., {Thorsett}, S.~E., {Golden}, A., \& {Goss}, W.~M. 2003, \bibinfo{title}{{The Distance and Radius of the Neutron Star PSR B0656+14},} \apjl, 593, L89, \dodoi{10.1086/378184}

\bibitem[{F. {Camilo} {et~al.}(2009){Camilo}, {Ransom}, {Gaensler}, \& {Lorimer}}]{2009ApJ...700L..34C}
{Camilo}, F., {Ransom}, S.~M., {Gaensler}, B.~M., \& {Lorimer}, D.~R. 2009, \bibinfo{title}{{Discovery of the Energetic Pulsar J1747-2809 in the Supernova Remnant G0.9+0.1},} \apjl, 700, L34, \dodoi{10.1088/0004-637X/700/1/L34}

\bibitem[{G.~L. {Case} \& D. {Bhattacharya}(1998){Case} \& {Bhattacharya}}]{1998ApJ...504..761C}
{Case}, G.~L., \& {Bhattacharya}, D. 1998, \bibinfo{title}{{A New {\ensuremath{\Sigma}}-D Relation and Its Application to the Galactic Supernova Remnant Distribution},} \apj, 504, 761, \dodoi{10.1086/306089}

\bibitem[{J.~L. {Caswell}(1985){Caswell}}]{1985AJ.....90.1224C}
{Caswell}, J.~L. 1985, \bibinfo{title}{{The supernova remnant G 54.5-0.3 and its environs.},} \aj, 90, 1224, \dodoi{10.1086/113829}

\bibitem[{J.~L. {Caswell} \& R.~F. {Haynes}(1975){Caswell} \& {Haynes}}]{1975MNRAS.173..649C}
{Caswell}, J.~L., \& {Haynes}, R.~F. 1975, \bibinfo{title}{{New 1612 MHz OH emission sources.},} \mnras, 173, 649, \dodoi{10.1093/mnras/173.3.649}

\bibitem[{J.~L. {Caswell} {et~al.}(2004){Caswell}, {McClure-Griffiths}, \& {Cheung}}]{2004MNRAS.352.1405C}
{Caswell}, J.~L., {McClure-Griffiths}, N.~M., \& {Cheung}, M.~C.~M. 2004, \bibinfo{title}{{Supernova remnant G292.2-0.5, its pulsar, and the Galactic magnetic field},} \mnras, 352, 1405, \dodoi{10.1111/j.1365-2966.2004.08030.x}

\bibitem[{J.~L. {Caswell} {et~al.}(1975){Caswell}, {Murray}, {Roger}, {Cole}, \& {Cooke}}]{1975A&A....45..239C}
{Caswell}, J.~L., {Murray}, J.~D., {Roger}, R.~S., {Cole}, D.~J., \& {Cooke}, D.~J. 1975, \bibinfo{title}{{Neutral hydrogen absorption measurements yielding kinematic distances for 42 continuum sources in the galactic plane},} \aap, 45, 239

\bibitem[{A.~N. {Cha} {et~al.}(1999){Cha}, {Sembach}, \& {Danks}}]{1999ApJ...515L..25C}
{Cha}, A.~N., {Sembach}, K.~R., \& {Danks}, A.~C. 1999, \bibinfo{title}{{The Distance to the Vela Supernova Remnant},} \apjl, 515, L25, \dodoi{10.1086/311968}

\bibitem[{B. {Chen} {et~al.}(2020){Chen}, {Wang}, {Hou}, {Yang}, {Li}, {Zhao}, \& {Jiang}}]{2020MNRAS.496.4637C}
{Chen}, B., {Wang}, S., {Hou}, L., {et~al.} 2020, \bibinfo{title}{{The distances to molecular clouds in the fourth Galactic quadrant},} \mnras, 496, 4637, \dodoi{10.1093/mnras/staa1827}

\bibitem[{B.~Q. {Chen} {et~al.}(2017){Chen}, {Liu}, {Ren}, {Yuan}, {Huang}, {Yu}, {Xiang}, {Wang}, {Tian}, \& {Zhang}}]{2017MNRAS.472.3924C}
{Chen}, B.~Q., {Liu}, X.~W., {Ren}, J.~J., {et~al.} 2017, \bibinfo{title}{{Mapping the three-dimensional dust extinction towards the supernova remnant S147 - the S147 dust cloud},} \mnras, 472, 3924, \dodoi{10.1093/mnras/stx2287}

\bibitem[{E. {Churchwell} {et~al.}(2009){Churchwell}, {Babler}, {Meade}, {Whitney}, {Benjamin}, {Indebetouw}, {Cyganowski}, {Robitaille}, {Povich}, {Watson}, \& {Bracker}}]{2009PASP..121..213C}
{Churchwell}, E., {Babler}, B.~L., {Meade}, M.~R., {et~al.} 2009, \bibinfo{title}{{The Spitzer/GLIMPSE Surveys: A New View of the Milky Way},} \pasp, 121, 213, \dodoi{10.1086/597811}

\bibitem[{J.~M. {Cordes} \& T.~J.~W. {Lazio}(2002){Cordes} \& {Lazio}}]{2002astro.ph..7156C}
{Cordes}, J.~M., \& {Lazio}, T.~J.~W. 2002, \bibinfo{title}{{NE2001.I. A New Model for the Galactic Distribution of Free Electrons and its Fluctuations},} arXiv e-prints, astro, \dodoi{10.48550/arXiv.astro-ph/0207156}

\bibitem[{D.~P. {Cox} {et~al.}(1999){Cox}, {Shelton}, {Maciejewski}, {Smith}, {Plewa}, {Pawl}, \& {R{\'o}{\.z}yczka}}]{1999ApJ...524..179C}
{Cox}, D.~P., {Shelton}, R.~L., {Maciejewski}, W., {et~al.} 1999, \bibinfo{title}{{Modeling W44 as a Supernova Remnant in a Density Gradient with a Partially Formed Dense Shell and Thermal Conduction in the Hot Interior. I. The Analytical Model},} \apj, 524, 179, \dodoi{10.1086/307781}

\bibitem[{J. {Deng} {et~al.}(2025){Deng}, {Wang}, {Jiang}, \& {Zhao}}]{2025ApJ...982...77D}
{Deng}, J., {Wang}, S., {Jiang}, B., \& {Zhao}, H. 2025, \bibinfo{title}{{The Multiwavelength Extinction Law and Its Variation in the Coalsack Molecular Cloud Based on the Gaia, APASS, SMSS, 2MASS, GLIMPSE, and WISE Surveys},} \apj, 982, 77, \dodoi{10.3847/1538-4357/adb431}

\bibitem[{J.~R. {Dickel}(1973){Dickel}}]{1973ApL....15...61D}
{Dickel}, J.~R. 1973, \bibinfo{title}{{The Distances to the Supernova Remnants IC443 and MSH11-61A},} \aplett, 15, 61

\bibitem[{B. {Din{\c{c}}el} {et~al.}(2015){Din{\c{c}}el}, {Neuh{\"a}user}, {Yerli}, {Ankay}, {Tetzlaff}, {Torres}, \& {Mugrauer}}]{2015MNRAS.448.3196D}
{Din{\c{c}}el}, B., {Neuh{\"a}user}, R., {Yerli}, S.~K., {et~al.} 2015, \bibinfo{title}{{Discovery of an OB runaway star inside SNR S147},} \mnras, 448, 3196, \dodoi{10.1093/mnras/stv124}

\bibitem[{R. {Dodson} {et~al.}(2003){Dodson}, {Legge}, {Reynolds}, \& {McCulloch}}]{2003ApJ...596.1137D}
{Dodson}, R., {Legge}, D., {Reynolds}, J.~E., \& {McCulloch}, P.~M. 2003, \bibinfo{title}{{The Vela Pulsar's Proper Motion and Parallax Derived from VLBI Observations},} \apj, 596, 1137, \dodoi{10.1086/378089}

\bibitem[{L.~N. {Driessen} {et~al.}(2018){Driessen}, {Dom{\v{c}}ek}, {Vink}, {Hessels}, {Arias}, \& {Gelfand}}]{2018ApJ...860..133D}
{Driessen}, L.~N., {Dom{\v{c}}ek}, V., {Vink}, J., {et~al.} 2018, \bibinfo{title}{{Investigating Galactic Supernova Remnant Candidates Using LOFAR},} \apj, 860, 133, \dodoi{10.3847/1538-4357/aac32e}

\bibitem[{G.~M. {Dubner} {et~al.}(1993){Dubner}, {Moffett}, {Goss}, \& {Winkler}}]{1993AJ....105.2251D}
{Dubner}, G.~M., {Moffett}, D.~A., {Goss}, W.~M., \& {Winkler}, P.~F. 1993, \bibinfo{title}{{Very Large Array Observations at 1465MHz of Galactic Supernova Remnants},} \aj, 105, 2251, \dodoi{10.1086/116603}

\bibitem[{G. {Ferrand} \& S. {Safi-Harb}(2012){Ferrand} \& {Safi-Harb}}]{2012AdSpR..49.1313F}
{Ferrand}, G., \& {Safi-Harb}, S. 2012, \bibinfo{title}{{A census of high-energy observations of Galactic supernova remnants},} Advances in Space Research, 49, 1313, \dodoi{10.1016/j.asr.2012.02.004}

\bibitem[{M. {Fich}(1986){Fich}}]{1986ApJ...303..465F}
{Fich}, M. 1986, \bibinfo{title}{{Supernova Remnants Associated with an H i ``Supershell'' in the Perseus Spiral Arm},} \apj, 303, 465, \dodoi{10.1086/164091}

\bibitem[{T. {Foster} \& C.~M. {Brunt}(2015){Foster} \& {Brunt}}]{2015AJ....150..147F}
{Foster}, T., \& {Brunt}, C.~M. 2015, \bibinfo{title}{{A CGPS Look at the Spiral Structure of the Outer Milky Way. I. Distances and Velocities to Star-forming Regions},} \aj, 150, 147, \dodoi{10.1088/0004-6256/150/5/147}

\bibitem[{D.~A. {Frail}(2011){Frail}}]{2011MmSAI..82..703F}
{Frail}, D.~A. 2011, \bibinfo{title}{{Supernova remnant shock - Molecular cloud interactions. Masers as tracers of hadronic particle acceleration},} \memsai, 82, 703, \dodoi{10.48550/arXiv.1108.4137}

\bibitem[{B.~M. {Gaensler} {et~al.}(1999){Gaensler}, {Brazier}, {Manchester}, {Johnston}, \& {Green}}]{1999MNRAS.305..724G}
{Gaensler}, B.~M., {Brazier}, K.~T.~S., {Manchester}, R.~N., {Johnston}, S., \& {Green}, A.~J. 1999, \bibinfo{title}{{SNR G320.4-01.2 and PSR B1509-58: new radio observations of a complex interacting system},} \mnras, 305, 724, \dodoi{10.1046/j.1365-8711.1999.02500.x}

\bibitem[{B.~M. {Gaensler} {et~al.}(1998){Gaensler}, {Manchester}, \& {Green}}]{1998MNRAS.296..813G}
{Gaensler}, B.~M., {Manchester}, R.~N., \& {Green}, A.~J. 1998, \bibinfo{title}{{Radio continuum and HI observations of the supernova remnant G296.8-00.3},} \mnras, 296, 813, \dodoi{10.1046/j.1365-8711.1998.01387.x}

\bibitem[{B.~M. {Gaensler} \& B.~J. {Wallace}(2003){Gaensler} \& {Wallace}}]{2003ApJ...594..326G}
{Gaensler}, B.~M., \& {Wallace}, B.~J. 2003, \bibinfo{title}{{A Multifrequency Radio Study of Supernova Remnant G292.0+1.8 and Its Pulsar Wind Nebula},} \apj, 594, 326, \dodoi{10.1086/376861}

\bibitem[{E. {Giacani} {et~al.}(2011){Giacani}, {Smith}, {Dubner}, \& {Loiseau}}]{2011AA...531A.138G}
{Giacani}, E., {Smith}, M.~J.~S., {Dubner}, G., \& {Loiseau}, N. 2011, \bibinfo{title}{{A new study of the supernova remnant G344.7-0.1 located in the vicinity of the unidentified TeV source HESS J1702-420},} \aap, 531, A138, \dodoi{10.1051/0004-6361/201116768}

\bibitem[{E.~B. {Giacani} {et~al.}(2000){Giacani}, {Dubner}, {Green}, {Goss}, \& {Gaensler}}]{2000AJ....119..281G}
{Giacani}, E.~B., {Dubner}, G.~M., {Green}, A.~J., {Goss}, W.~M., \& {Gaensler}, B.~M. 2000, \bibinfo{title}{{The Interstellar Matter in the Direction of the Supernova Remnant G296.5+10.0 and the Central X-Ray Source 1E 1207.4-5209},} \aj, 119, 281, \dodoi{10.1086/301173}

\bibitem[{D.~A. {Green}(2009){Green}}]{2009MNRAS.399..177G}
{Green}, D.~A. 2009, \bibinfo{title}{{Re-identification of G35.6-0.4 as a supernova remnant},} \mnras, 399, 177, \dodoi{10.1111/j.1365-2966.2009.14957.x}

\bibitem[{D.~A. {Green}(2019){Green}}]{2019JApA...40...36G}
{Green}, D.~A. 2019, \bibinfo{title}{{A revised catalogue of 294 Galactic supernova remnants},} Journal of Astrophysics and Astronomy, 40, 36, \dodoi{10.1007/s12036-019-9601-6}

\bibitem[{J.~W. {Hewitt} \& F. {Yusef-Zadeh}(2009){Hewitt} \& {Yusef-Zadeh}}]{2009ApJ...694L..16H}
{Hewitt}, J.~W., \& {Yusef-Zadeh}, F. 2009, \bibinfo{title}{{Discovery of New Interacting Supernova Remnants in the Inner Galaxy},} \apjl, 694, L16, \dodoi{10.1088/0004-637X/694/1/L16}

\bibitem[{A.~P. {Igoshev} {et~al.}(2022){Igoshev}, {Frantsuzova}, {Gourgouliatos}, {Tsichli}, {Konstantinou}, \& {Popov}}]{2022MNRAS.514.4606I}
{Igoshev}, A.~P., {Frantsuzova}, A., {Gourgouliatos}, K.~N., {et~al.} 2022, \bibinfo{title}{{Initial periods and magnetic fields of neutron stars},} \mnras, 514, 4606, \dodoi{10.1093/mnras/stac1648}

\bibitem[{N. {Junkes} {et~al.}(1992){Junkes}, {Fuerst}, \& {Reich}}]{1992A&AS...96....1J}
{Junkes}, N., {Fuerst}, E., \& {Reich}, W. 1992, \bibinfo{title}{{G 54.4-0.3 : CO shell and star formation region surrounding a shell-type supernova remnant. I. Properties of the CO shell.},} \aaps, 96, 1

\bibitem[{V.~M. {Kaspi}(1998){Kaspi}}]{1998nspt.conf..401K}
{Kaspi}, V.~M. 1998, in Neutron Stars and Pulsars: Thirty Years after the Discovery, ed. N.~{Shibazaki}, 401, \dodoi{10.48550/arXiv.astro-ph/9803026}

\bibitem[{A. {Khalatyan} {et~al.}(2024){Khalatyan}, {Anders}, {Chiappini}, {Queiroz}, {Nepal}, {dal Ponte}, {Jordi}, {Guiglion}, {Valentini}, {Torralba Elipe}, {Steinmetz}, {Pantaleoni-Gonz{\'a}lez}, {Malhotra}, {Jim{\'e}nez-Arranz}, {Enke}, {Casamiquela}, \& {Ard{\`e}vol}}]{2024A&A...691A..98K}
{Khalatyan}, A., {Anders}, F., {Chiappini}, C., {et~al.} 2024, \bibinfo{title}{{Transferring spectroscopic stellar labels to 217 million Gaia DR3 XP stars with SHBoost},} \aap, 691, A98, \dodoi{10.1051/0004-6361/202451427}

\bibitem[{C.~D. {Kilpatrick} {et~al.}(2016){Kilpatrick}, {Bieging}, \& {Rieke}}]{2016ApJ...816....1K}
{Kilpatrick}, C.~D., {Bieging}, J.~H., \& {Rieke}, G.~H. 2016, \bibinfo{title}{{A Systematic Survey for Broadened CO Emission toward Galactic Supernova Remnants},} \apj, 816, 1, \dodoi{10.3847/0004-637X/816/1/1}

\bibitem[{C.~S. {Kochanek} {et~al.}(2024){Kochanek}, {Raymond}, \& {Caldwell}}]{2024ApJ...968...94K}
{Kochanek}, C.~S., {Raymond}, J.~C., \& {Caldwell}, N. 2024, \bibinfo{title}{{The Distance to the S147 Supernova Remnant},} \apj, 968, 94, \dodoi{10.3847/1538-4357/ad4493}

\bibitem[{R. {Kothes}(2013){Kothes}}]{2013AA...560A..18K}
{Kothes}, R. 2013, \bibinfo{title}{{Distance and age of the pulsar wind nebula 3C 58},} \aap, 560, A18, \dodoi{10.1051/0004-6361/201219839}

\bibitem[{R. {Kothes} \& S.~M. {Dougherty}(2007){Kothes} \& {Dougherty}}]{2007AA...468..993K}
{Kothes}, R., \& {Dougherty}, S.~M. 2007, \bibinfo{title}{{The distance and neutral environment of the massive stellar cluster Westerlund 1},} \aap, 468, 993, \dodoi{10.1051/0004-6361:20077309}

\bibitem[{R. {Kothes} {et~al.}(2001){Kothes}, {Uyaniker}, \& {Pineault}}]{2001ApJ...560..236K}
{Kothes}, R., {Uyaniker}, B., \& {Pineault}, S. 2001, \bibinfo{title}{{The Supernova Remnant G106.3+2.7 and Its Pulsar-Wind Nebula: Relics of Triggered Star Formation in a Complex Environment},} \apj, 560, 236, \dodoi{10.1086/322511}

\bibitem[{T.~L. {Landecker} {et~al.}(1987){Landecker}, {Vaneldik}, {Dewdney}, \& {Routledge}}]{1987AJ.....94..111L}
{Landecker}, T.~L., {Vaneldik}, J.~F., {Dewdney}, P.~E., \& {Routledge}, D. 1987, \bibinfo{title}{{Observations at 408 MHz of the Supernova Remnant HB3 (G132.6 + 1.5)},} \aj, 94, 111, \dodoi{10.1086/114453}

\bibitem[{J.~M. {Lattimer} \& M. {Prakash}(2004){Lattimer} \& {Prakash}}]{2004Sci...304..536L}
{Lattimer}, J.~M., \& {Prakash}, M. 2004, \bibinfo{title}{{The Physics of Neutron Stars},} Science, 304, 536, \dodoi{10.1126/science.1090720}

\bibitem[{D. {Leahy} \& W. {Tian}(2010){Leahy} \& {Tian}}]{2010ASPC..438..365L}
{Leahy}, D., \& {Tian}, W. 2010, in Astronomical Society of the Pacific Conference Series, Vol. 438, The Dynamic Interstellar Medium: A Celebration of the Canadian Galactic Plane Survey, ed. R.~{Kothes}, T.~L. {Landecker}, \& A.~G. {Willis}, 365

\bibitem[{D.~A. {Leahy} {et~al.}(2013){Leahy}, {Green}, \& {Ranasinghe}}]{2013MNRAS.436..968L}
{Leahy}, D.~A., {Green}, K., \& {Ranasinghe}, S. 2013, \bibinfo{title}{{X-ray and radio observations of the {\ensuremath{\gamma}} Cygni supernova remnant G78.2+2.1},} \mnras, 436, 968, \dodoi{10.1093/mnras/stt1596}

\bibitem[{D.~A. {Leahy} \& S. {Ranasinghe}(2012){Leahy} \& {Ranasinghe}}]{2012MNRAS.423..718L}
{Leahy}, D.~A., \& {Ranasinghe}, S. 2012, \bibinfo{title}{{Radio observations of CTB80: detection of the snowplough in an old supernova remnant},} \mnras, 423, 718, \dodoi{10.1111/j.1365-2966.2012.20909.x}

\bibitem[{D.~A. {Leahy} {et~al.}(2008){Leahy}, {Tian}, \& {Wang}}]{2008AJ....136.1477L}
{Leahy}, D.~A., {Tian}, W., \& {Wang}, Q.~D. 2008, \bibinfo{title}{{Distance Determination to the Crab-Like Pulsar Wind Nebula G54.1+0.3 and the Search for its Supernova Remnant Shell},} \aj, 136, 1477, \dodoi{10.1088/0004-6256/136/4/1477}

\bibitem[{D.~A. {Leahy} \& W.~W. {Tian}(2007){Leahy} \& {Tian}}]{2007AA...461.1013L}
{Leahy}, D.~A., \& {Tian}, W.~W. 2007, \bibinfo{title}{{Radio spectrum and distance of the SNR HB9},} \aap, 461, 1013, \dodoi{10.1051/0004-6361:20065895}

\bibitem[{D.~A. {Leahy} \& W.~W. {Tian}(2008){Leahy} \& {Tian}}]{2008AJ....135..167L}
{Leahy}, D.~A., \& {Tian}, W.~W. 2008, \bibinfo{title}{{The Distances of SNR W41 and Overlapping H II Regions},} \aj, 135, 167, \dodoi{10.1088/0004-6256/135/1/167}

\bibitem[{L. {Li} {et~al.}(2023){Li}, {Wang}, {Chen}, \& {Jiang}}]{2023ApJ...956...26L}
{Li}, L., {Wang}, S., {Chen}, X., \& {Jiang}, Q. 2023, \bibinfo{title}{{The Ultraviolet to Mid-infrared Extinction Law of the Taurus Molecular Cloud Based on the Gaia DR3, GALEX, APASS, Pan-STARRS1, 2MASS, and WISE Surveys},} \apj, 956, 26, \dodoi{10.3847/1538-4357/aced8a}

\bibitem[{L. {Lindegren} {et~al.}(2021){Lindegren}, {Bastian}, {Biermann}, {Bombrun}, {de Torres}, {Gerlach}, {Geyer}, {Hern{\'a}ndez}, {Hilger}, {Hobbs}, {Klioner}, {Lammers}, {McMillan}, {Ramos-Lerate}, {Steidelm{\"u}ller}, {Stephenson}, \& {van Leeuwen}}]{2021A&A...649A...4L}
{Lindegren}, L., {Bastian}, U., {Biermann}, M., {et~al.} 2021, \bibinfo{title}{{Gaia Early Data Release 3. Parallax bias versus magnitude, colour, and position},} \aap, 649, A4, \dodoi{10.1051/0004-6361/202039653}

\bibitem[{D.~S. {Mathewson} \& J.~N. {Clarke}(1973){Mathewson} \& {Clarke}}]{1973ApJ...180..725M}
{Mathewson}, D.~S., \& {Clarke}, J.~N. 1973, \bibinfo{title}{{Supernova remnants in the Large Magellanic Cloud.},} \apj, 180, 725, \dodoi{10.1086/152002}

\bibitem[{V.~J. {Napoli} {et~al.}(2011){Napoli}, {McSwain}, {Marsh Boyer}, \& {Roettenbacher}}]{2011PASP..123.1262N}
{Napoli}, V.~J., {McSwain}, M.~V., {Marsh Boyer}, A.~N., \& {Roettenbacher}, R.~M. 2011, \bibinfo{title}{{The Distance of the {\ensuremath{\gamma}}-Ray Binary 1FGL J1018.6-5856},} \pasp, 123, 1262, \dodoi{10.1086/662692}

\bibitem[{M.~Z. {Pavlovi{\'c}} {et~al.}(2013){Pavlovi{\'c}}, {Uro{\v{s}}evi{\'c}}, {Vukoti{\'c}}, {Arbutina}, \& {G{\"o}ker}}]{2013ApJS..204....4P}
{Pavlovi{\'c}}, M.~Z., {Uro{\v{s}}evi{\'c}}, D., {Vukoti{\'c}}, B., {Arbutina}, B., \& {G{\"o}ker}, {\"U}.~D. 2013, \bibinfo{title}{{The Radio Surface-brightness-to-Diameter Relation for Galactic Supernova Remnants: Sample Selection and Robust Analysis with Various Fitting Offsets},} \apjs, 204, 4, \dodoi{10.1088/0067-0049/204/1/4}

\bibitem[{S. {Pineault} {et~al.}(1993){Pineault}, {Landecker}, {Madore}, \& {Gaumont-Guay}}]{1993AJ....105.1060P}
{Pineault}, S., {Landecker}, T.~L., {Madore}, B., \& {Gaumont-Guay}, S. 1993, \bibinfo{title}{{The Supernova Remnant CTA1 and the Surrounding Interstellar Medium},} \aj, 105, 1060, \dodoi{10.1086/116493}

\bibitem[{J.~A. {Pons} \& D. {Vigan{\`o}}(2019){Pons} \& {Vigan{\`o}}}]{2019LRCA....5....3P}
{Pons}, J.~A., \& {Vigan{\`o}}, D. 2019, \bibinfo{title}{{Magnetic, thermal and rotational evolution of isolated neutron stars},} Living Reviews in Computational Astrophysics, 5, 3, \dodoi{10.1007/s41115-019-0006-7}

\bibitem[{D.~C. {Price} {et~al.}(2021){Price}, {Flynn}, \& {Deller}}]{2021PASA...38...38P}
{Price}, D.~C., {Flynn}, C., \& {Deller}, A. 2021, \bibinfo{title}{{A comparison of Galactic electron density models using PyGEDM},} \pasa, 38, e038, \dodoi{10.1017/pasa.2021.33}

\bibitem[{V. {Radhakrishnan} {et~al.}(1972){Radhakrishnan}, {Goss}, {Murray}, \& {Brooks}}]{1972ApJS...24...49R}
{Radhakrishnan}, V., {Goss}, W.~M., {Murray}, J.~D., \& {Brooks}, J.~W. 1972, \bibinfo{title}{{The Parkes Survey of 21-CENTIMETER Absorption in Discrete-Source Spectra. III. 21- Centimeter Absorption Measurements on 41 Galactic Sources North of Declination -48 degrees},} \apjs, 24, 49, \dodoi{10.1086/190249}

\bibitem[{S. {Ranasinghe} \& D. {Leahy}(2022){Ranasinghe} \& {Leahy}}]{2022ApJ...940...63R}
{Ranasinghe}, S., \& {Leahy}, D. 2022, \bibinfo{title}{{Distances, Radial Distribution, and Total Number of Galactic Supernova Remnants},} \apj, 940, 63, \dodoi{10.3847/1538-4357/ac940a}

\bibitem[{S. {Ranasinghe} \& D. {Leahy}(2023){Ranasinghe} \& {Leahy}}]{2023ApJS..265...53R}
{Ranasinghe}, S., \& {Leahy}, D. 2023, \bibinfo{title}{{A Statistical Analysis of Galactic Radio Supernova Remnants},} \apjs, 265, 53, \dodoi{10.3847/1538-4365/acc1de}

\bibitem[{S. {Ranasinghe} \& D.~A. {Leahy}(2017){Ranasinghe} \& {Leahy}}]{2017ApJ...843..119R}
{Ranasinghe}, S., \& {Leahy}, D.~A. 2017, \bibinfo{title}{{Distances to Supernova Remnants G31.9+0.0 and G54.4-0.3 Associated with Molecular Clouds},} \apj, 843, 119, \dodoi{10.3847/1538-4357/aa7894}

\bibitem[{S. {Ranasinghe} \& D.~A. {Leahy}(2018){Ranasinghe} \& {Leahy}}]{2018AJ....155..204R}
{Ranasinghe}, S., \& {Leahy}, D.~A. 2018, \bibinfo{title}{{Revised Distances to 21 Supernova Remnants},} \aj, 155, 204, \dodoi{10.3847/1538-3881/aab9be}

\bibitem[{W. {Reich} \& E. {Braunsfurth}(1981){Reich} \& {Braunsfurth}}]{1981AA....99...17R}
{Reich}, W., \& {Braunsfurth}, E. 1981, \bibinfo{title}{{2.7 GHz observations of the three old supernova remnants, CTB 1, G 116.5+1.1, and G 114.3+0.3 with the Effelsberg 100-m telescope.},} \aap, 99, 17

\bibitem[{E.~M. {Reynoso} {et~al.}(2017){Reynoso}, {Cichowolski}, \& {Walsh}}]{2017MNRAS.464.3029R}
{Reynoso}, E.~M., {Cichowolski}, S., \& {Walsh}, A.~J. 2017, \bibinfo{title}{{A high-resolution H I study towards the supernova remnant Puppis A and its environments},} \mnras, 464, 3029, \dodoi{10.1093/mnras/stw2219}

\bibitem[{E.~M. {Reynoso} {et~al.}(2003){Reynoso}, {Green}, {Johnston}, {Dubner}, {Giacani}, \& {Goss}}]{2003MNRAS.345..671R}
{Reynoso}, E.~M., {Green}, A.~J., {Johnston}, S., {et~al.} 2003, \bibinfo{title}{{Observations of the neutral hydrogen surrounding the radio-quiet neutron star RX J0822-4300 in Puppis A},} \mnras, 345, 671, \dodoi{10.1046/j.1365-8711.2003.06978.x}

\bibitem[{E.~M. {Reynoso} {et~al.}(2004){Reynoso}, {Green}, {Johnston}, {Goss}, {Dubner}, \& {Giacani}}]{2004PASA...21...82R}
{Reynoso}, E.~M., {Green}, A.~J., {Johnston}, S., {et~al.} 2004, \bibinfo{title}{{Influence of the Neutron Star 1E 161348-5055 in RCW 103 on the Surrounding Medium},} \pasa, 21, 82, \dodoi{10.1071/AS03053}

\bibitem[{E.~M. {Reynoso} {et~al.}(2006){Reynoso}, {Johnston}, {Green}, \& {Koribalski}}]{2006MNRAS.369..416R}
{Reynoso}, E.~M., {Johnston}, S., {Green}, A.~J., \& {Koribalski}, B.~S. 2006, \bibinfo{title}{{High-resolution HI and radio continuum observations of the SNR G290.1-0.8},} \mnras, 369, 416, \dodoi{10.1111/j.1365-2966.2006.10325.x}

\bibitem[{M.~S.~E. {Roberts} \& C.~L. {Brogan}(2008){Roberts} \& {Brogan}}]{2008ApJ...681..320R}
{Roberts}, M. S.~E., \& {Brogan}, C.~L. 2008, \bibinfo{title}{{Discovery of a New X-Ray Filled Radio Supernova Remnant around the Pulsar Wind Nebula in 3EG J1809-2328},} \apj, 681, 320, \dodoi{10.1086/588419}

\bibitem[{M.~T. {Ruiz}(1983){Ruiz}}]{1983AJ.....88.1210R}
{Ruiz}, M.~T. 1983, \bibinfo{title}{{Spectrophotometry of the optical emission from RCW 103 and Milne 23.},} \aj, 88, 1210, \dodoi{10.1086/113411}

\bibitem[{M.~T. {Ruiz} \& J. {May}(1986){Ruiz} \& {May}}]{1986ApJ...309..667R}
{Ruiz}, M.~T., \& {May}, J. 1986, \bibinfo{title}{{MSH 10-53: A Supernova Remnant Interacting with Molecular Clouds},} \apj, 309, 667, \dodoi{10.1086/164634}

\bibitem[{M. {S{\'a}nchez-Cruces} {et~al.}(2018){S{\'a}nchez-Cruces}, {Rosado}, {Fuentes-Carrera}, \& {Ambrocio-Cruz}}]{2018MNRAS.473.1705S}
{S{\'a}nchez-Cruces}, M., {Rosado}, M., {Fuentes-Carrera}, I., \& {Ambrocio-Cruz}, P. 2018, \bibinfo{title}{{Kinematics of the Galactic Supernova Remnant G109.1-1.0 (CTB 109)},} \mnras, 473, 1705, \dodoi{10.1093/mnras/stx2460}

\bibitem[{F.~D. {Seward}(1985){Seward}}]{1985ComAp..11...15S}
{Seward}, F.~D. 1985, \bibinfo{title}{{Supernova remnants containing neutron stars.},} Comments on Astrophysics, 11, 15

\bibitem[{S.-S. {Shan} {et~al.}(2019){Shan}, {Zhu}, {Tian}, {Zhang}, {Yang}, \& {Zhang}}]{2019RAA....19...92S}
{Shan}, S.-S., {Zhu}, H., {Tian}, W.-W., {et~al.} 2019, \bibinfo{title}{{The distance measurements of supernova remnants in the fourth Galactic quadrant},} Research in Astronomy and Astrophysics, 19, 092, \dodoi{10.1088/1674-4527/19/7/92}

\bibitem[{S.~S. {Shan} {et~al.}(2018){Shan}, {Zhu}, {Tian}, {Zhang}, {Zhang}, {Wu}, \& {Yang}}]{2018ApJS..238...35S}
{Shan}, S.~S., {Zhu}, H., {Tian}, W.~W., {et~al.} 2018, \bibinfo{title}{{Distances of Galactic Supernova Remnants Using Red Clump Stars},} \apjs, 238, 35, \dodoi{10.3847/1538-4365/aae07a}

\bibitem[{Y. {Su} {et~al.}(2011){Su}, {Chen}, {Yang}, {Koo}, {Zhou}, {Lu}, {Jeong}, \& {DeLaney}}]{2011ApJ...727...43S}
{Su}, Y., {Chen}, Y., {Yang}, J., {et~al.} 2011, \bibinfo{title}{{Molecular Environment and Thermal X-ray Spectroscopy of the Semicircular Young Composite Supernova Remnant 3C 396},} \apj, 727, 43, \dodoi{10.1088/0004-637X/727/1/43}

\bibitem[{L. {Supan} {et~al.}(2016){Supan}, {Supanitsky}, \& {Castelletti}}]{2016AA...589A..51S}
{Supan}, L., {Supanitsky}, A.~D., \& {Castelletti}, G. 2016, \bibinfo{title}{{The environment of the {\ensuremath{\gamma}}-ray emitting SNR G338.3-0.0: a hadronic interpretation for HESS J1640-465},} \aap, 589, A51, \dodoi{10.1051/0004-6361/201527962}

\bibitem[{H. {Suzuki} {et~al.}(2020){Suzuki}, {Bamba}, {Enokiya}, {Yamaguchi}, {Plucinsky}, \& {Odaka}}]{2020ApJ...893..147S}
{Suzuki}, H., {Bamba}, A., {Enokiya}, R., {et~al.} 2020, \bibinfo{title}{{Uniform Distribution of the Extremely Overionized Plasma Associated with the Supernova Remnant G359.1-0.5},} \apj, 893, 147, \dodoi{10.3847/1538-4357/ab80ba}

\bibitem[{W.~W. {Tian} {et~al.}(2007){Tian}, {Haverkorn}, \& {Zhang}}]{2007MNRAS.378.1283T}
{Tian}, W.~W., {Haverkorn}, M., \& {Zhang}, H.~Y. 2007, \bibinfo{title}{{Characteristics of the supernova remnant G351.7+0.8 and a distance argument against its association with PSR J1721-3532},} \mnras, 378, 1283, \dodoi{10.1111/j.1365-2966.2007.11613.x}

\bibitem[{W.~W. {Tian} \& D.~A. {Leahy}(2006){Tian} \& {Leahy}}]{2006AA...455.1053T}
{Tian}, W.~W., \& {Leahy}, D.~A. 2006, \bibinfo{title}{{The radio SNR G65.1+0.6 and its associated pulsar J1957+2831},} \aap, 455, 1053, \dodoi{10.1051/0004-6361:20065140}

\bibitem[{W.~W. {Tian} \& D.~A. {Leahy}(2012){Tian} \& {Leahy}}]{2012MNRAS.421.2593T}
{Tian}, W.~W., \& {Leahy}, D.~A. 2012, \bibinfo{title}{{Distances of the TeV supernova remnant complex CTB 37 towards the Galactic bar},} \mnras, 421, 2593, \dodoi{10.1111/j.1365-2966.2012.20491.x}

\bibitem[{A.~G.~G.~M. {Tielens}(2005){Tielens}}]{2005pcim.book.....T}
{Tielens}, A.~G.~G.~M. 2005, {The Physics and Chemistry of the Interstellar Medium}

\bibitem[{P.~F. {Vel{\'a}zquez} {et~al.}(2002){Vel{\'a}zquez}, {Dubner}, {Goss}, \& {Green}}]{2002AJ....124.2145V}
{Vel{\'a}zquez}, P.~F., {Dubner}, G.~M., {Goss}, W.~M., \& {Green}, A.~J. 2002, \bibinfo{title}{{Investigation of the Large-scale Neutral Hydrogen near the Supernova Remnant W28},} \aj, 124, 2145, \dodoi{10.1086/342936}

\bibitem[{J.~P.~W. {Verbiest} {et~al.}(2012){Verbiest}, {Weisberg}, {Chael}, {Lee}, \& {Lorimer}}]{2012ApJ...755...39V}
{Verbiest}, J.~P.~W., {Weisberg}, J.~M., {Chael}, A.~A., {Lee}, K.~J., \& {Lorimer}, D.~R. 2012, \bibinfo{title}{{On Pulsar Distance Measurements and Their Uncertainties},} \apj, 755, 39, \dodoi{10.1088/0004-637X/755/1/39}

\bibitem[{J. {Vink}(2004){Vink}}]{2004ApJ...604..693V}
{Vink}, J. 2004, \bibinfo{title}{{Revealing the Obscured Supernova Remnant Kesteven 32 with Chandra},} \apj, 604, 693, \dodoi{10.1086/381930}

\bibitem[{J. {Vink}(2008){Vink}}]{2008ApJ...689..231V}
{Vink}, J. 2008, \bibinfo{title}{{The Kinematics of Kepler's Supernova Remnant as Revealed by Chandra},} \apj, 689, 231, \dodoi{10.1086/592375}

\bibitem[{S. {Wang} \& X. {Chen}(2019){Wang} \& {Chen}}]{2019ApJ...877..116W}
{Wang}, S., \& {Chen}, X. 2019, \bibinfo{title}{{The Optical to Mid-infrared Extinction Law Based on the APOGEE, Gaia DR2, Pan-STARRS1, SDSS, APASS, 2MASS, and WISE Surveys},} \apj, 877, 116, \dodoi{10.3847/1538-4357/ab1c61}

\bibitem[{S. {Wang} \& X. {Chen}(2023){Wang} \& {Chen}}]{2023ApJ...946...43W}
{Wang}, S., \& {Chen}, X. 2023, \bibinfo{title}{{The Optical to Infrared Extinction Law of Magellanic Clouds Based on Red Supergiants and Classical Cepheids},} \apj, 946, 43, \dodoi{10.3847/1538-4357/acb647}

\bibitem[{S. {Wang} \& B.~W. {Jiang}(2014){Wang} \& {Jiang}}]{2014ApJ...788L..12W}
{Wang}, S., \& {Jiang}, B.~W. 2014, \bibinfo{title}{{Universality of the Near-infrared Extinction Law Based on the APOGEE Survey},} \apjl, 788, L12, \dodoi{10.1088/2041-8205/788/1/L12}

\bibitem[{S. {Wang} {et~al.}(2020){Wang}, {Zhang}, {Jiang}, {Zhao}, {Chen}, {Chen}, {Gao}, \& {Liu}}]{2020AA...639A..72W}
{Wang}, S., {Zhang}, C., {Jiang}, B., {et~al.} 2020, \bibinfo{title}{{Distances to the supernova remnants in the inner disk},} \aap, 639, A72, \dodoi{10.1051/0004-6361/201936868}

\bibitem[{J.~M. {Yao} {et~al.}(2017){Yao}, {Manchester}, \& {Wang}}]{2017ApJ...835...29Y}
{Yao}, J.~M., {Manchester}, R.~N., \& {Wang}, N. 2017, \bibinfo{title}{{A New Electron-density Model for Estimation of Pulsar and FRB Distances},} \apj, 835, 29, \dodoi{10.3847/1538-4357/835/1/29}

\bibitem[{Y. {Yao} {et~al.}(2024){Yao}, {Ji}, {Koposov}, \& {Limberg}}]{2024MNRAS.52710937Y}
{Yao}, Y., {Ji}, A.~P., {Koposov}, S.~E., \& {Limberg}, G. 2024, \bibinfo{title}{{200 000 candidate very metal-poor stars in Gaia DR3 XP spectra},} \mnras, 527, 10937, \dodoi{10.1093/mnras/stad3775}

\bibitem[{A. {Yar-Uyaniker} {et~al.}(2004){Yar-Uyaniker}, {Uyaniker}, \& {Kothes}}]{2004ApJ...616..247Y}
{Yar-Uyaniker}, A., {Uyaniker}, B., \& {Kothes}, R. 2004, \bibinfo{title}{{Distance of Three Supernova Remnants from H I Line Observations in a Complex Region: G114.3+0.3, G116.5+1.1, and CTB 1 (G116.9+0.2)},} \apj, 616, 247, \dodoi{10.1086/424794}

\bibitem[{B. {Yu} {et~al.}(2019){Yu}, {Chen}, {Jiang}, \& {Zijlstra}}]{2019MNRAS.488.3129Y}
{Yu}, B., {Chen}, B.~Q., {Jiang}, B.~W., \& {Zijlstra}, A. 2019, \bibinfo{title}{{Three-dimensional dust mapping of 12 supernovae remnants in the Galactic anticentre},} \mnras, 488, 3129, \dodoi{10.1093/mnras/stz1940}

\bibitem[{X. {Zhang} {et~al.}(2023){Zhang}, {Green}, \& {Rix}}]{2023MNRAS.524.1855Z}
{Zhang}, X., {Green}, G.~M., \& {Rix}, H.-W. 2023, \bibinfo{title}{{Parameters of 220 million stars from Gaia BP/RP spectra},} \mnras, 524, 1855, \dodoi{10.1093/mnras/stad1941}

\bibitem[{H. {Zhao} {et~al.}(2018){Zhao}, {Jiang}, {Gao}, {Li}, \& {Sun}}]{2018ApJ...855...12Z}
{Zhao}, H., {Jiang}, B., {Gao}, S., {Li}, J., \& {Sun}, M. 2018, \bibinfo{title}{{The Distance to and the Near-infrared Extinction of the Monoceros Supernova Remnant},} \apj, 855, 12, \dodoi{10.3847/1538-4357/aaacd0}

\bibitem[{H. {Zhao} {et~al.}(2020){Zhao}, {Jiang}, {Li}, {Chen}, {Yu}, \& {Wang}}]{2020ApJ...891..137Z}
{Zhao}, H., {Jiang}, B., {Li}, J., {et~al.} 2020, \bibinfo{title}{{A Systematic Study of the Dust of Galactic Supernova Remnants. I. The Distance and the Extinction},} \apj, 891, 137, \dodoi{10.3847/1538-4357/ab75ef}

\bibitem[{H. {Zhao} {et~al.}(2024){Zhao}, {Wang}, {Jiang}, {Li}, {Fan}, {Ren}, \& {Ma}}]{2024ApJ...974..138Z}
{Zhao}, H., {Wang}, S., {Jiang}, B., {et~al.} 2024, \bibinfo{title}{{Data-driven Stellar Intrinsic Colors and Dust Reddenings for Spectrophotometric Data: From the Blue-edge Method to a Machine Learning Approach},} \apj, 974, 138, \dodoi{10.3847/1538-4357/ad6d64}

\bibitem[{P. {Zhou} {et~al.}(2020){Zhou}, {Zhou}, {Chen}, {Wang}, {Vink}, \& {Wang}}]{2020ApJ...905...99Z}
{Zhou}, P., {Zhou}, X., {Chen}, Y., {et~al.} 2020, \bibinfo{title}{{Revisiting the Distance, Environment, and Supernova Properties of SNR G57.2+0.8 that Hosts SGR 1935+2154},} \apj, 905, 99, \dodoi{10.3847/1538-4357/abc34a}

\bibitem[{X. {Zhou} {et~al.}(2016){Zhou}, {Yang}, {Fang}, {Su}, {Sun}, \& {Chen}}]{2016ApJ...833....4Z}
{Zhou}, X., {Yang}, J., {Fang}, M., {et~al.} 2016, \bibinfo{title}{{Interaction between the Supernova Remnant HB 3 and the Nearby Star-forming Region W3},} \apj, 833, 4, \dodoi{10.3847/0004-637X/833/1/4}

\bibitem[{X. {Zhou} {et~al.}(2023){Zhou}, {Su}, {Yang}, {Chen}, {Sun}, {Jiang}, {Wang}, {Wang}, {Zhang}, {Xu}, {Yan}, {Yuan}, {Chen}, {Ao}, \& {Ma}}]{2023ApJS..268...61Z}
{Zhou}, X., {Su}, Y., {Yang}, J., {et~al.} 2023, \bibinfo{title}{{A Systematic Study of Associations between Supernova Remnants and Molecular Clouds},} \apjs, 268, 61, \dodoi{10.3847/1538-4365/acee7f}

\bibitem[{H. {Zhu} {et~al.}(2013){Zhu}, {Tian}, {Torres}, {Pedaletti}, \& {Su}}]{2013ApJ...775...95Z}
{Zhu}, H., {Tian}, W.~W., {Torres}, D.~F., {Pedaletti}, G., \& {Su}, H.~Q. 2013, \bibinfo{title}{{A Kinematic Distance Study of the Planetary Nebulae-Supernova remnant-H II Region Complex at G35.6-0.5},} \apj, 775, 95, \dodoi{10.1088/0004-637X/775/2/95}

\end{thebibliography}

\bibliographystyle{aasjournalv7}



\end{document}